\documentclass[aps,prd,onecolumn,nofootinbib,groupedaddress,superscriptaddress]{revtex4}  
\usepackage{graphicx}
\usepackage{epstopdf}
\usepackage{amsmath}
\usepackage{amsfonts}
\usepackage{amssymb}
\usepackage{appendix}
\usepackage{enumerate}
\usepackage{comment}
\usepackage{bbold}
\usepackage[shortlabels]{enumitem}
\usepackage{color}
\usepackage{slashed}
\usepackage{subfigure}
\usepackage{setspace}
\usepackage{footnote}
\usepackage{multirow}
\usepackage{longtable}
\usepackage[colorlinks = true,
            linkcolor = blue,
            urlcolor  = blue,
            citecolor = blue,
            anchorcolor = blue]{hyperref}
\usepackage[capitalize]{cleveref}
\usepackage{braket}
\usepackage[compat=1.1.0]{tikz-feynman}
\usepackage{multirow}
 \usepackage{relsize}
\usepackage{wasysym}

\usepackage[a4paper, portrait, margin=1in]{geometry}

\newcommand{\beq}{\begin{equation}}
\newcommand{\eeq}{\end{equation}}

\newcommand{\newtext}[1]{{\color{black} #1}}

\begin{document}

\singlespacing

{\hfill FERMILAB-PUB-21-362-T, NUHEP-TH/21-04}

\title{Energy-Dependent Neutrino Mixing Parameters at Oscillation Experiments}

\author{K.~S.~Babu}
\affiliation{Department of Physics, Oklahoma State University, Stillwater, OK, 74078, USA}
\author{Vedran Brdar} 
\affiliation{Northwestern University, Department of Physics \& Astronomy, 2145 Sheridan Road, Evanston, IL 60208, USA}
\affiliation{Theoretical Physics Department, Fermilab, P.O. Box 500, Batavia, IL 60510, USA}
\author{Andr\'{e} de Gouv\^{e}a} 
\affiliation{Northwestern University, Department of Physics \& Astronomy, 2145 Sheridan Road, Evanston, IL 60208, USA}
\author{Pedro A.~N.~Machado}
\affiliation{Theoretical Physics Department, Fermilab, P.O. Box 500, Batavia, IL 60510, USA}

\begin{abstract}
%%%%%%%%%%%%%%
Neutrino mixing parameters 
are subject to quantum corrections and hence are scale dependent. This means that the mixing parameters associated to the production and detection of neutrinos need not coincide since these processes are characterized by different energy scales. We show that, in the presence of relatively light new physics, the scale dependence of the mixing parameters can lead to observable consequences in long-baseline neutrino oscillation experiments, such as T2K and NOvA, and in neutrino telescopes like IceCube.  We discuss some of the experimental signatures of this scenario, including zero-baseline flavor transitions, new sources of CP-invariance violation, and apparent inconsistencies among measurements of mixing angles at different experiments or oscillation channels. Finally, we present simple, ultraviolet-complete models of neutrino masses which lead to observable running of the neutrino mixing matrix below the weak scale.

%%%%%%%%%%%%%%%%%
\end{abstract}

\maketitle

%%%%%%%%%%%%%%%%%%
\section{Introduction}
\label{sec:intro}
\setcounter{equation}{0}
%%%%%%%%%%%%%%%%%% 
\noindent
The discovery of neutrino oscillations towards the end of the last century  \cite{Fukuda:1998mi,Ahmad:2002jz} launched a diverse, world-wide experimental neutrino oscillation program that is expected to continue, at least, well into the next decade with the DUNE \cite{Abi:2020wmh} and Hyper-Kamiokande \cite{Abe:2018uyc} projects, currently under construction. It aims at measuring, sometimes with exquisite precision, the neutrino oscillation phenomenon via a variety of oscillation channels, baselines, and experimental conditions. The ultimate goal is to, broadly speaking, test the three-massive-active-neutrinos paradigm that postulates the existence of three neutral leptons with different masses that interact only via the neutral-current and charged-current weak interactions, as prescribed by the Standard Model of particle physics (SM). 

While the three-massive-active-neutrinos paradigm provides an excellent fit to virtually all neutrino data,\footnote{There are experimental results that do not fit the three-massive-active-neutrinos paradigm including searches for electron (anti)neutrino appearance at the LSND \cite{Athanassopoulos:1996jb} and MiniBooNE \cite{AguilarArevalo:2010wv,Aguilar-Arevalo:2018gpe,Aguilar-Arevalo:2020nvw} experiments as well as the so-called reactor antineutrino anomaly \cite{Mueller:2011nm,Huber:2011wv}. Explanations to these remain elusive and will not be considered in any detail here.} the current data allow for the presence of more new physics in the neutrino sector. Different, well-motivated new physics scenarios can be probed by the current and next generation of neutrino experiments. These include the existence of new, light neutral-fermion degrees of freedom that mix with the active neutrinos (``sterile neutrinos''), and new neutrino--matter interactions that manifest themselves at the energies of interest via four-fermion operators (``non-standard interactions (NSI)''). The latter are usually associated to heavy new physics and are in general strongly constrained, in the absence of a fair amount of fine-tuning, by charged-lepton processes  \cite{Davidson:2003ha,Ibarra:2004pe,Gavela:2008ra,Biggio:2009kv, Davidson:2019iqh}, with some exceptions, see for example Refs.~\cite{Farzan:2015hkd,Farzan:2016wym,Babu:2017olk,Farzan:2017xzy}.

Here, we explore potential new phenomena associated to new, relatively light degrees of freedom that interact almost exclusively with neutrinos. In these scenarios, constraints from the charged-lepton sector are significantly weaker while other constraints, including those associated to the existence of new, light degrees of freedom in the early universe, can be avoided. New interactions between neutrinos and new, light particles can impact neutrino experiments in two different ways: (i) the new states can be produced when neutrinos are produced or detected, leading to changes in the kinematics and flavor structure of neutrino scattering (see, for example, \cite{Berryman:2018ogk,Brdar:2018qqj,Chang:2020jwl,Hurtado:2020vlj,Abdallah:2020biq,Brdar:2020tle} and references therein) or (ii) quantum corrections associated to the virtual exchange of the new degrees of freedom can modify neutrino production and detection. Here, we concentrate on the latter which, to the best of our knowledge, has not been explored extensively in the literature.

At the core of the idea is the fact that, once higher-order corrections are included, the parameters that describe neutrino oscillations are energy dependent. This is neither new nor surprising. For example, the renormalization-group (RG) running of neutrino oscillation parameters between the neutrino-mass-generating scale (often assumed to be much higher than the scale of electroweak symmetry breaking) and the weak scale has been the subject of intense investigation in the literature, see for instance Refs.~\cite{Babu:1993qv,Chankowski:1993tx,Antusch:2001ck,Casas:1999tg,Balaji:2000au,Antusch:2003kp,Antusch:2005gp,Goswami:2009yy}. 
The running of mass and mixing parameters is not exclusive to the neutrino sector: quark masses run significantly above the GeV scale and the running of the bottom and the top masses has been directly observed at lepton and hadron colliders~\cite{Abreu:1997ey, Brandenburg:1999nb, Abbiendi:2001tw, Abdallah:2005cv, Langenfeld:2009wd, Sirunyan:2019jyn}. The CKM  matrix elements are also expected to run above the weak scale \cite{Babu:1987im}; this running, however, has never been observed. Below the weak scale, these matrix elements can be treated as constant in the absence of light new physics. This is usually a good approximation since new, light degrees of freedom that couple to quarks are strongly constrained.

Different from the quark sector, however, new light degrees of freedom that couple predominantly to neutrinos are not strongly constrained. If these exist, RG running effects may be relevant in the context of neutrino oscillation experiments.
In this paper, we show that neutrino oscillation probabilities are affected through the mismatch between the leptonic mixing matrix evaluated at the scale (or more precisely momentum transfer) corresponding to the neutrino production and the one at which neutrinos are detected. A careful treatment of the oscillation phenomenon, therefore, requires -- for a fixed neutrino energy! -- twice as many relevant mixing angles (production and detection values). The number of CP-odd parameters is also larger. We find that while running effects are already strongly constrained, they can impact significantly the current and next generation of neutrino oscillation experiments, including T2K and NOvA. Their presence may lead, for example, to apparent inconsistencies between measurements of oscillation parameters between T2K and NOvA and between ``accelerator'' and ``reactor'' measurements of oscillation parameters. On the other hand, there are simple new physics scenarios that lead to significant low-energy running of the oscillation parameters, including some that are related to the origin of nonzero neutrino masses. Hence, these effects are not only possible in principle, they may be accessible if neutrino masses are a consequence of relatively light, new physics. 

This manuscript is organized as follows. In \cref{sec:running}, we set the stage by introducing and discussing the general idea. 
\cref{sec:impact} contains a detailed treatment of neutrino oscillation probabilities in the presence of RG-running effects. First, in \cref{subsec:vacuum}, we compute the vacuum oscillation probability in general and discuss the more familiar, less cumbersome two-flavor case in some detail; in this simplified framework, we discuss the oscillation probabilities in different useful limits including circumstances when the running effects are small and the case in which the oscillation baseline is zero. We discuss the more useful but much more cumbersome three-flavor scenario concentrating on subsets of the parameter space. In \cref{subsec:matter}, we discuss matter effects, which are relevant for the long baseline experiments under consideration. For the two-flavor case, exact expressions are given while the realistic three-flavor case can only be tackled, for all pragmatic purposes, numerically. In \cref{sec:models}, we discuss two concrete models that lead to large RG-running effects. A quantitative study of the consequences of these models is presented in \cref{sec:results}. There, we discuss some consequences for T2K and NOvA in light of constraints from short-baseline experiments (\cref{subsec:long}). In \cref{subsec:IC}, we scrutinize the impact on the flavor composition of ultra-high-energy neutrinos. Finally, we conclude in \cref{sec:conclusion}.

%%%%%%%%%%%%%%%%%%
\section{Running of Neutrino Mixing Parameters}
\label{sec:running}
\setcounter{equation}{0}
%%%%%%%%%%%%%%%%%
\noindent
A simple way to understand that quantum corrections can lead to nontrivial effects in neutrino oscillations is to investigate the charged-current weak interactions in the mass basis for both charged leptons and neutrinos. In more detail
\begin{equation}
-{\cal L}\supset \frac{g}{\sqrt{2}}U_{\alpha i} \bar{\ell}_{\alpha}\slashed{W}^-P_L\nu_i + H.c.~,
\end{equation}
where $\ell_{\alpha}$, $\alpha=e,\mu,\tau$, are the charged-lepton fields, $\nu_i$, $i=1,2,3$, are the neutrino fields with well defined masses $m_1,m_2,m_3$, respectively, $g$ is the $SU(2)_L$ gauge coupling, $P_L$ is the left-chiral projection operator, and $U_{\alpha i}$ are the elements of the leptonic mixing matrix. The product $(gU_{\alpha i})$ can be interpreted as the coupling between a $W$-boson, a charged lepton $\ell_\alpha$, and a neutrino $\nu_i$. Once higher-order quantum effects are included, the question of interest here is whether these allow $(gU_{\alpha i})$ to change relative to one another. When this happens, as we discuss carefully below, we can say that the mixing matrix ``runs.''

\begin{figure}
\centering
	\includegraphics[scale=0.9]{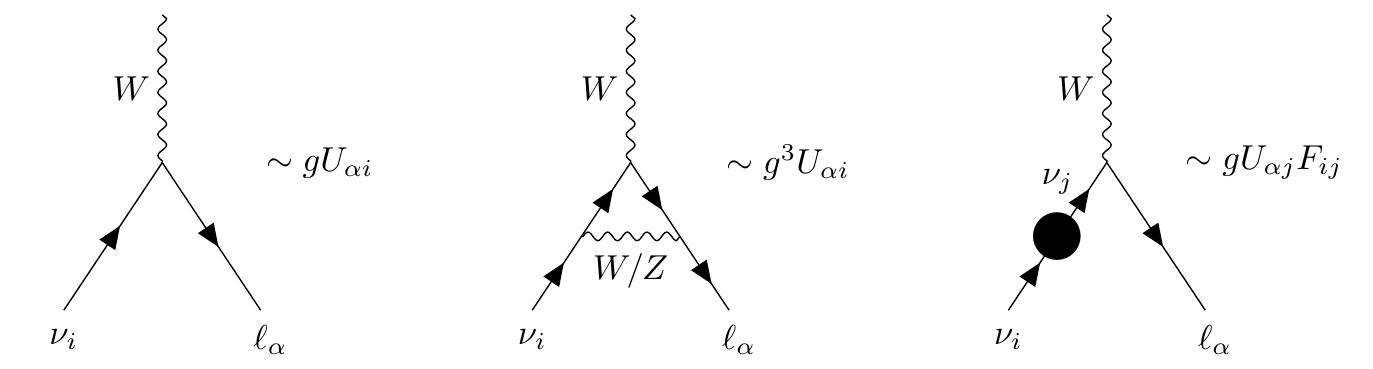} 
	\caption{Left: Leading order Feynman diagram for the $W$-boson, charged lepton $\ell_\alpha$, neutrino $\nu_i$ coupling. Center: Sample one-loop electroweak correction. This contribution is proportional to $(gU_{\alpha j})(gU^*_{\beta j})(gU_{\beta i})=g^3U_{\alpha i}$ in the limit where the fermion masses are negligible and $U$ is unitary. Right: Sample one-loop correction from a new interaction that modifies the neutrino propagator. This contribution is proportional to $(gU_{\alpha j})F_{ij}$.}
	\label{fig:feyn1}
\end{figure}

It is easy to see that, ignoring fermion Yukawa coupling effects, higher order electroweak corrections to $(gU_{\alpha i})$ are trivially proportional to $(gU_{\alpha i})$: $(gU_{\alpha i})\to (gU_{\alpha i})\times F$, where $F$ does not depend on the indices $\alpha$ or $i$. Fig.~\ref{fig:feyn1} (center-panel) depicts one of the many higher-order one-loop electroweak corrections, for illustrative purposes. The presence of new interactions changes the picture significantly as long as these have a nontrivial flavor structure. For example, a new interaction that involves only neutrinos and new degrees of freedom, depicted schematically in Fig.~\ref{fig:feyn1} (right-panel), will modify the neutrino propagator and, in turn, modify $(gU_{\alpha i})\to (gU_{\alpha i})+\sum_j(gU_{\alpha j})\times F_{ij}$. If the ``loop-factors'' $F_{ij}$ depend on $i,j=1,2,3$, the $(gU_{\alpha i})$ change in a flavor-dependent way. 

A simple, concrete model that would manifest itself in this way is adding a gauge-singlet scalar $\Phi$ to the SM field content and allowing for neutrino--scalar Yukawa interaction of the form $h_{ij}\Phi\bar{\nu}_i\nu_j$ (after electroweak symmetry breaking). Starting at the one-loop level, it is easy to see how the dark circle in Fig.~\ref{fig:feyn1} (right-panel) is realized. This is not the model we explore here. Instead, we will concentrate on two ultraviolet complete models,\footnote{The gauge-invariant realization of the $\Phi\bar{\nu}\nu$ operators is not renormalizable.} introduced and discussed in more detail in Sec.~\ref{sec:models}.

Following the renormalization-group approach to capture the finite quantum corrections, when the dust settles, we assume that we can replace $(gU_{\alpha i})\to (g(Q^2)U_{\alpha i}(Q^2))$ in a way that both $g$ and $U$ depend on the momentum-transfer (squared) $Q^2$ that characterizes the interaction: both gauge coupling constant and the elements of the mixing matrix ``run.'' Note that we are assuming that, for a fixed value of $Q^2$, $U_{\alpha i}(Q^2)$ can always be expressed as elements of a unitary matrix so it is meaningful to discuss $U(Q^2)$ as a running mixing matrix. When discussing neutrino production or detection, it is often convenient to define the neutrino flavor eigenstates $\nu_{\alpha}$. The discussion above implies that, given the existence of the new, generation-dependent interactions, the neutrino flavor eigenstates are energy-scale dependent.  

Neutrino flavor change as a function of the distance between source and detector depends on which linear combinations of $\nu_i$ couple to the different charged leptons for both the production and detection processes -- the flavor eigenstates mentioned above -- and on the differences of the squares of the neutrino masses. While quantum corrections also lead to running masses, in neutrino oscillations one is interested in the pole masses.\footnote{The running mass should be used in evaluating the neutrino production and detection processes. Neutrino masses, however, are small enough that their impact is negligible in virtually all processes of interest.} The reason is we are interested in neutrinos that propagate a macroscopic distance. In the language of quantum field theory, the only contribution to the amplitude that characterizes neutrino production plus detection comes from on-shell neutrino exchange: the contributions from virtual neutrino exchange, to exquisite precision, cancel out. 
\newtext{As a consequence of neutrinos being on-shell, production and detection of neutrinos, which are associated to different energy scales, can be treated separately \cite{Volobuev:2017izt}.} 

In summary, RG effects impact neutrino oscillations in the sense that they render the neutrino mixing matrix $Q^2$-dependent. We discuss flavor-change in this context in great detail in the next section and return to concrete, phenomenologically-safe models that lead to nontrivial $U(Q^2)$ in Sec.~\ref{sec:models}.

%%%%%%%%%%%%%%%%%%
\section{Neutrino Oscillation Phenomenology}
\label{sec:impact}
\setcounter{equation}{0}
%%%%%%%%%%%%%%%%%%
\noindent
As discussed above, the new interactions of interest imply that the neutrino charged-current-weak-interaction eigenstates $\nu_e,\nu_{\mu},\nu_{\tau}$ depend on the scale  of the neutrino-production and neutrino-detection processes. We can choose the scale of the couplings to be the so-called Lorentz-invariant ``momentum transfer'' $Q^2$: if we adopt all momenta in a vertex to be incoming, then $Q^2\equiv|(p_\nu+p_\ell)^2|$, where $p_\nu$ and $p_\ell$ are, respectively, the momenta of the neutrino and charged lepton involved in the charged-current process of interest. 
For example, in $\pi\to\mu\nu_\mu$ we have $Q^2=m_\pi^2$.
In general, in a physics process characterized by $Q^2$ where a charged lepton $\ell_\alpha$ ($\alpha=e,\mu,\tau$) is absorbed and a neutrino is produced, the coherent linear combination of neutrino mass eigenstates $\nu_i$ ($i=1,2,3$) produced is 
\begin{equation}
\nu_{\alpha}(Q^2)=U_{\alpha i}(Q^2)\nu_i. 
\end{equation}
We are interested in neutrinos that propagate a finite distance so, as discussed in the previous section, when referring to neutrino mass eigenstates, we always refer to on-shell mass eigenstates with on-shell masses $m_i$, that is $\nu_i\equiv \nu_i(Q^2=m_{\nu_i}^2)$. Note that we assume neutrinos are always produced and detected via the charged-current weak interactions. 

In vacuum, a neutrino mass eigenstate with energy $E$ is an eigenstate of the propagation Hamiltonian $H$.
Its evolution, assuming that the neutrino is ultra-relativistic, has the familiar form
\begin{equation}
|\nu_i(L)\rangle  = \exp[-im_i^2L/2E]|\nu_i(0)\rangle,
\end{equation}
where $L$ is the distance propagated or the baseline. 
The Dirac bracket notation refers to the flavor wave-function of the neutrino state (e.g., a three-dimensional Hilbert space assuming there are three independent neutrino states). The probability $P_{\alpha\beta}$ that a neutrino is produced associated to a charged lepton $\ell_{\alpha}$ in a process characterized by $Q^2_p$ and detected, a distance $L$ away, in association with a charged lepton $\ell_{\beta}$ in a process characterized by  $Q^2_d$ can be trivially, but carefully, computed:
\begin{align}
P_{\alpha\beta}(L) = \left|\langle\nu_{\beta}(Q^2_d)|\exp[-iHL]|\nu_{\alpha}(Q^2_p)\rangle\right|^2= 
\left|\sum_i U_{\beta i}(Q^2_d)U^*_{\alpha i}(Q^2_p)\exp[-im_i^2L/2E]\right|^2. 
\label{eq:Pab}
\end{align}

If the kinematics of the production and detection processes are the same, we obtain the familiar vacuum neutrino oscillation expressions. In general, however, this is not the case. Imagine an experimental setup where neutrinos are produced in charged-pion decay: $\pi^+\to \mu^+\nu$. In this case, $Q^2_p=m_{\pi}^2$, independent from the pion and neutrino laboratory energies. Further imagine that the neutrino is detected via $\nu+n\to e^-+p$. In the neutron's rest-frame, for large enough neutrino energies, $Q^2_d\sim Em_n$; $Q^2_d$ depends on the neutrino detector-frame energy and, clearly, has no relation to the pion mass. 
In the subsections that follow, we explore the consequences and subtleties of Eq.~(\ref{eq:Pab}) and discuss how to handle the propagation of neutrinos through matter when the mixing matrix is $Q^2$ dependent. 

Before proceeding, we wish to highlight that, for a fixed value of $Q^2_p$ and $Q^2_d$, the oscillation formalism we will explore here is similar to what one would obtain when considering the hypothesis that there is new physics in neutrino production and detection, and that the impact of new physics is process and  flavor dependent \cite{Grossman:1995wx} (for a more recent discussion, see also Ref.~\cite{Falkowski:2019xoe}).  The two scenarios, however, are not identical since, in the setup under consideration here, there are different mixing matrixes for different neutrino-scattering energies, even if the physics processes in question are the same; our $U_{\alpha}$ depend on the momentum transfer, not the nature of the neutrino-production and neutrino-detection processes. As an aside, our discussion here is a little more complete relative to the one on Ref.~\cite{Grossman:1995wx} as we look into three-flavor effects, matter effects, and new CP-violating phenomena more carefully.

\subsection{Vacuum oscillations: two and more flavors}
\label{subsec:vacuum}

We first consider the simplified case of two charged-leptons and two neutrinos -- $e$, $\mu$, $\nu_1$, and $\nu_2$ for concreteness -- propagating in vacuum. The most general two-by-two $Q^2$-dependent mixing matrix can be expressed as 
\begin{equation}
U(Q^2)=
\left(\begin{array}{cc} 1 & 0 \\ 0 &  e^{i\gamma(Q^2)}\end{array}\right)
\left(\begin{array}{cc} \cos\theta(Q^2) & \sin\theta(Q^2) \\ -\sin\theta(Q^2) & \cos\theta(Q^2)\end{array}\right)\left(\begin{array}{cc} e^{i\tilde\alpha(Q^2)} & 0 \\ 0 &  e^{i\tilde \beta(Q^2)}\end{array}\right),
\end{equation}
where we indicate the $Q^2$ dependency of the mixing parameters 
$\gamma,\theta,\tilde\alpha,\tilde\beta$ 
explictly. Not all these parameters are physical. We can redefine, with impunity, the kets  $|\nu_e(Q^2)\rangle\to \exp[i\zeta(Q^2)]|\nu_e(Q^2)\rangle$ and $|\nu_\mu(Q^2)\rangle\to\exp[i\eta(Q^2)]|\nu_\mu(Q^2)\rangle$, for every value of $Q^2$, such that two among three complex phases can be removed. For example, 
\begin{equation}
U(Q^2)=\left(\begin{array}{cc} \cos\theta(Q^2) & \sin\theta(Q^2) \\ -\sin\theta(Q^2) & \cos\theta(Q^2)\end{array}\right)\left(\begin{array}{cc} 1 & 0 \\ 0 &  e^{i\tilde\beta(Q^2)}\end{array}\right),
\label{eq:U2flavors}
\end{equation}
allows one to access the answer to all possible oscillation-related questions.  Finally, we are allowed to also phase-redefine the mass-eigenstate kets $|\nu_i\rangle$ with impunity. Hence, we can choose $\tilde\beta(Q^2)$ in Eq.~(\ref{eq:U2flavors}) to vanish {\sl at some fixed value of $Q^2$}. In the most general case, therefore, the oscillation probabilities in Eq.~(\ref{eq:Pab}) will depend on one mass-squared difference 
$\Delta m^2\equiv m_2^2-m_1^2$ and three mixing parameters: 
\begin{equation}
  \theta(Q^2_p)\equiv \theta_p, \quad \theta(Q^2_d)\equiv\theta_d, \quad {\rm and}\quad \tilde\beta(Q^2_d)-\tilde\beta(Q^2_p)\equiv\beta.
\end{equation}
Throughout, in order to unambiguously define the mass-eigenstates, we choose $m_2>m_1$. 

A different subtlety lies in the physical range for the mixing parameters. $Q^2$-dependent sign redefinitions of $|\nu_e\rangle$ and $|\nu_\mu\rangle$ allow one to choose, for example, $\theta(Q^2)\in[-\pi/2,\pi/2]$. Sign redefinitions of the mass eingenstates allow one to constrain $\theta$ to a specific quadrant (for example, $\theta(Q^2)\in[0,\pi/2]$) {\sl for some fixed value of $Q^2$}. Hence, if we choose $\theta_p$ to lie in the first quadrant, there is no guarantee that the same will be true of $\theta_d$.

Before proceeding, we wish to point out that the phase $\tilde\beta(Q^2)$ in Eq.~(\ref{eq:U2flavors}) bears a strong resemblance to the so-called Majorana phase. However, they are not the same and should not be confused. Majorana phases are only physical when neutrinos are Majorana fermions and only manifest themselves in phenomena related, directly or indirectly, to lepton-number violation. The relative phase $\beta$ manifests itself in ordinary flavor-conversions and has nothing to do with lepton-number violation. It is physical for both Majorana and Dirac neutrinos. 
It is fair to ask about the origin and interpretation of this CP-odd new physics parameter and why it impacts neutrino oscillations even when there are only two families of leptons. One way to understand it is as follows. With two families and assuming the massive neutrinos are Dirac fermions, the weak-interaction couplings and the fermion masses can be chosen real. In the model under consideration here, however, there are more flavor-dependent interactions. If there are CP-violating couplings in the new physics sector, those will manifest themselves at higher-order in charged-current processes and can mediate CP-violating effects. In oscillation-language, these are parameterized by the (running) CP-odd phase $\beta$ introduced above.
Incidentally, it is trivial to show using Eqs.~(\ref{eq:Pab}) and (\ref{eq:U2flavors}) that the effect of $\beta$ in the oscillation probabilities in vacuum is to ``shift'' the oscillation phase: $\Delta m^2 L/2E\to \Delta m^2 L/2E + \beta$.

With all this in mind, for two-flavors, Eq.~(\ref{eq:Pab}) can be expressed as
\begin{equation}
P_{e\mu} = P_{\mu e} = \sin^2(\theta_p-\theta_d) + \sin2\theta_p\sin2\theta_d\sin^2\left(\frac{\Delta m^2L}{4E}+\frac{\beta}{2}\right),
\label{eq:Pemu}
\end{equation}
and 
\begin{equation}
P_{ee} = P_{\mu\mu} =\cos^2(\theta_p-\theta_d) - \sin2\theta_p\sin2\theta_d\sin^2\left(\frac{\Delta m^2L}{4E}+\frac{\beta}{2}\right).
\label{eq:Pee}
\end{equation}
The equality of $P_{\mu e}$ and $P_{e\mu}$ and $P_{ee}$ and $P_{\mu\mu}$ is a consequence of the unitary evolution of the neutrino states and the fact that $|\nu_{e}(Q^2)\rangle$ and  $|\nu_{\mu}(Q^2)\rangle$, for any $Q^2$, are a complete basis for the two-dimensional Hilbert space.\footnote{In the two neutrino framework, one can show that $P_{ee}+P_{e\mu}=1$ and $P_{ee}+P_{\mu e}=1$. These translate into $P_{\mu e}=P_{e \mu}$.} This is independent from the presence of the complex relative phase $\beta$. However, $P_{\mu e}=P_{e\mu}$ does not mean that $T$-invariance is guaranteed. More carefully, Eq.~(\ref{eq:Pemu}) states that $P(\nu_e(Q^2_p)\to \nu_{\mu}(Q^2_d))=P(\nu_\mu(Q^2_p)\to \nu_e(Q^2_d))$. $T$-invariance is the statement $P(\nu_e(Q^2_p)\to \nu_{\mu}(Q^2_d))=P(\nu_\mu(Q^2_d)\to \nu_e(Q^2_p))$. $P(\nu_\mu(Q^2_d)\to \nu_e(Q^2_p))$ is given by Eq.~(\ref{eq:Pemu}) with $\theta_d\leftrightarrow\theta_p$ and $\beta\to -\beta$ so $T$-invariance is violated if $\beta\neq 0,\pi$.

For antineutrinos, $P_{\bar{\alpha}\bar{\beta}}(\theta_p,\theta_d,\beta)=P_{\alpha\beta}(\theta_p,\theta_d,-\beta)$ so $CP$-invariance is also violated if $\beta\neq 0,\pi$. Note that, in general, $P_{\alpha\alpha}\neq P_{\bar{\alpha}\bar{\alpha}}$, which could be interpreted as an \textit{apparent} violation of CPT.
However, this does not signal violation of the CPT-theorem since CPT-invariance implies $P(\nu_\alpha(Q^2_p)\to \nu_\alpha(Q^2_d))=P(\bar{\nu}_\alpha(Q^2_d)\to \bar{\nu}_\alpha(Q^2_p))$, which is satisfied.

For $Q^2_d=Q^2_p$, we recover the well-known two-flavor oscillation-expressions. In general, however, the situation is qualitatively different. For example, in the limit $L\to 0$, flavor is violated if either $\theta_{p}\neq\theta_d$ or $\beta\neq 0$. This is easy to understand. In either case, the mixing matrices are different at different values of $Q^2$ so the linear combinations of neutrinos that couple to $e$ and $\mu$ are different: $\langle\nu_\alpha(Q^2_d)|\nu_\beta(Q^2_p\rangle\neq \delta_{\alpha\beta}$. These zero baseline effects, in practice, will constrain running effects to be relatively small, as we discuss more quantitatively in Sec.~\ref{sec:results}. The complex relative phase $\beta$, in turn, leads to a phase shift in the oscillatory phenomenon. 
This leads, for example, to different behaviors of the oscillation probabilities at zero and small baselines.
At zero baseline, a small $\beta$ still induces an oscillation phase and thus $P_{\alpha \gamma}=\delta_{\alpha \gamma}+\mathcal{O}(\beta^2)$.
At small-but-finite baseline $L$, while the standard oscillation probability goes as $P_{\alpha \gamma}^{\rm std}\sim \delta_{\alpha \gamma}+ \mathcal{O}(\Delta_{ij}^2)$, a nonzero $\beta$ would induce $P_{\alpha \gamma}\sim \mathcal{O}(\beta\Delta_{ij})$, where we have defined $\Delta_{ij}\equiv \Delta m^2_{ij}L/2E$.

In the limit where the running effects are small, it is easier to appreciate analytically the impact of the new physics effects. Assuming $\theta_{d}-\theta_{p}=\epsilon_\theta+{\cal O}(\epsilon_\theta^2)$ and $\beta=\epsilon_\beta+{\cal O}(\epsilon_\beta^2)$, both $\epsilon_\theta,\epsilon_\beta\ll 1$ and unrelated to one another, 
\begin{equation}
P_{e\mu} = P_{\mu e} = \epsilon_\theta^2 +  {\cal O}(\epsilon_\theta^4)+\left[\sin^22\theta_d-\sin4\theta_d\epsilon_\theta+{\cal O}(\epsilon_\theta^2)\right]\left[\sin^2\left(\frac{\Delta m^2L}{4E}\right)+\frac{\epsilon_\beta}{2}\sin\left(\frac{\Delta m^2L}{2E}\right)+{\cal O}(\epsilon_\beta^2)\right].
\label{eq:Pemuapprox}
\end{equation}
and 
\begin{equation}
P_{ee} = P_{\mu\mu} = 1-\epsilon_\theta^2 +  {\cal O}(\epsilon_\theta^4)-\left(\sin^22\theta_d-\sin4\theta_d\epsilon_\theta+{\cal O}(\epsilon_\theta^2)\right)\left[\sin^2\left(\frac{\Delta m^2L}{4E}\right)+\frac{\epsilon_\beta}{2}\sin\left(\frac{\Delta m^2L}{2E}\right)+{\cal O}(\epsilon_\beta^2)\right].
\end{equation}
In the zero-baseline limit, the new-physics effects are ${\cal O}(\epsilon_\theta^2,\epsilon_\beta^2)$, quadratically small in the limit $\epsilon_\theta,\epsilon_\beta\ll 1$. For a finite baseline, instead, the new-physics effects are ${\cal O}(\epsilon_\theta,\epsilon_\beta)$. For example,
\begin{equation}
P_{e\mu} = P_{\mu e} = \left(\sin^22\theta_d
-\epsilon_\theta \sin4\theta_d\right)\sin^2\left(\frac{\Delta m^2L}{4E}\right) +
\frac{\epsilon_\beta}{2}\sin^22\theta_d\sin\left(\frac{\Delta m^2L}{2E}\right) + {\cal O}(\epsilon_\theta^2,\epsilon_\beta^2,\epsilon_\theta\epsilon_\beta).
\label{eq:Pemuapprox2}
\end{equation}
If the effects of mixing-angle running are not large, long-baseline experiments are, in some sense, more sensitive than short-baseline experiments. 

The case of three charged-leptons and neutrinos -- $e$, $\mu$, $\tau$, $\nu_1$, $\nu_2$, $\nu_3$ -- is straightforward but more cumbersome. Taking advantage of the invariance of observables on the overall phases of $|\nu_{\alpha}(Q^2)\rangle$, $\alpha=e,\mu,\tau$ and $|\nu_{i}\rangle$, $i=1,2,3$, the most general $3\times 3$ $Q^2$-dependent mixing matrix can be parameterized as 
\begin{align}
U(Q^2) = &\left(\begin{array}{ccc} 1 & 0 & 0 \\ 0 & c_{23}(Q^2) & s_{23}(Q^2) \\ 0 & -s_{23}(Q^2) & c_{23}(Q^2) \end{array}\right)
\left(\begin{array}{ccc} c_{13}(Q^2) & 0 & s_{13}(Q^2)e^{-i\delta(Q^2)} \\ 0 & 1 & 0 \\ -s_{13}(Q^2)e^{i\delta(Q^2)} & 0 & c_{13}(Q^2) \end{array}\right)\\
&\hspace{3cm}\times \left(\begin{array}{ccc} c_{12}(Q^2) & s_{12}(Q^2) & 0 \\ -s_{12}(Q^2) & c_{12}(Q^2) & 0 \\ 0 & 0 & 1 \end{array}\right)
 \left(\begin{array}{ccc} 1 & 0 & 0 \\ 0 & e^{i\tilde\alpha(Q^2)} & 0 \\ 0 & 0 & e^{i\tilde\beta(Q^2)} \end{array}\right),
 \label{eq:U3}
\end{align}
where $c_{ij}(Q^2),s_{ij}(Q^2)$ are short-hand for $\cos\theta_{ij}(Q^2),\sin\theta_{ij}(Q^2)$, respectively, for $ij=12,13,23$. The complex phases $\tilde\alpha(Q^2)$ and $\tilde\beta(Q^2)$ can be chosen such that they vanish at some value of $Q^2$. 

The vacuum oscillation probabilities will be given by  Eq.~(\ref{eq:Pab}) with the elements of the mixing matrix as defined in Eq.~(\ref{eq:U3}). These will depend on the usual mass-squared differences $\Delta m^2_{31}\equiv m_3^2-m_1^2$ and $\Delta m^2_{21}=m_2^2-m_1^2$ (the third mass-squared difference is not independent, $\Delta m^2_{32}\equiv m_3^2-m_2^2=\Delta m^2_{31}-\Delta m^2_{21}$), six mixing angles $\theta_{ij}(Q_p^2),\theta_{ij}(Q_d^2)$, $ij=12,13,23$, two ``Dirac'' phases $\delta(Q^2_p),\delta(Q^2_d)$ and two additional complex-phase-differences, $\alpha\equiv\tilde\alpha(Q^2_d)-\tilde\alpha(Q^2_p)$ and $\beta\equiv\tilde\beta(Q^2_d)-\tilde\beta(Q^2_p)$. 
As in the two neutrino case, $\alpha$ and $\beta$ will induce a shift in the solar and atmospheric oscillation phases: $\Delta_{21}\to \Delta_{21}+\alpha$, $\Delta_{31}\to \Delta_{31}+\beta$, $\Delta_{32}\to \Delta_{32}+\beta-\alpha$.
The mass eigenstates can be unambiguously defined in a variety of ways. Here, it pays off to adopt a definition that does not depend on the mixing matrix in order to avoid a $Q^2$-dependent definition. Concretely, we take the standard definition: $m_2^2>m_1^2$ and $|\Delta m^2_{31}|>\Delta m^2_{21}$. $\Delta m^2_{31}>0$ defines the `normal' mass ordering (NO), $\Delta m^2_{31}<0$ the `inverted' one (IO).\footnote{A different choice would have been $|U_{e1}(Q^2)|^2>|U_{e2}(Q^2)|^2>|U_{e3}(Q^2)|^2$. This clearly depends on $Q^2$ and may lead to confusion. There is nothing wrong, however, with using something like $|U_{e1}(Q^2=0)|^2>|U_{e2}(Q^2=0)|^2>|U_{e3}(Q^2=0)|^2$.}

The expressions for the oscillation probabilities for three flavors in the $Q^2$-dependent case are lengthy. We have nevertheless found several features in the zero-baseline limit and in the limit of small RG effects that turn out to be instructive. 
Let us first turn to the expressions in the zero baseline ($L=0$) limit. 
For simplicity, we take $\theta_{13}=0$ and $\theta_{23}=\pi/4$ at production and assume that the difference between various parameters at different scales is small. We define $\epsilon_{ij}\equiv\theta_{ij}(Q^2_d)-\theta_{ij}(Q^2_p)$, $\epsilon_{\delta}=\delta(Q^2_d)-\delta(Q^2_p)$, $\epsilon_{\alpha}=\alpha$, and $\epsilon_{\beta}=\beta$. Note that $|\alpha|,|\beta|\ll 2\pi$. Hence, $\epsilon_{ij}$ is the amount the angle runs between momentum-transfers corresponding to production and detection.
We omit the ``production'' subscript and thus it should be understood that all angles in the expressions below correspond to the momentum scale of neutrino production. Finally, below, we keep up to quadratic terms in all $\epsilon$s.
This leads to the zero-baseline survival probabilities
\begin{align}
P_{ee}\simeq\bigg|1-\frac{\epsilon_{12}^2+\epsilon_{13}^2}{2}-\frac{\epsilon_\alpha^2}{2} s_{12}^2
	+ i\epsilon_\alpha s_{12}^2\bigg|^2 \simeq
1-\epsilon_{12}^2-\epsilon_{13}^2-\frac{1}{4} \epsilon_\alpha^2 \sin^2 2\theta_{12},
\label{eq:zeroe}
\end{align}
and
\begin{align}
P_{\mu\mu}&=\bigg|1-\frac14\left(\epsilon_{12}^2+\epsilon_{13}^2+2 \epsilon_{23}^2 +\epsilon_\alpha^2 c_{12}^2 +\epsilon_\beta^2 - \epsilon_{13}\epsilon_\alpha s_\delta \sin2\theta_{12} + 2\epsilon_{12}\epsilon_{13}c_\delta\right)
+\frac i2\left(\epsilon_\alpha c^2_{12} + \epsilon_\beta\right)   \bigg|^2   \nonumber \\
 & \simeq
1- \frac{1}{4} (\epsilon_\beta-\epsilon_\alpha c^2_{12})^2 - \epsilon_{23}^2 
	- \frac12\left(\epsilon_\alpha c_{12}s_{12}-\epsilon_{13}s_\delta\right)^2-\frac12\left(\epsilon_{12}+\epsilon_{13}c_\delta\right)^2,
	\label{eq:zeromu}
\end{align}
where $c_\delta=\cos\delta$ and $s_\delta=\sin\delta$.

On the other hand, zero-baseline appearance probabilities include
\begin{align}
P_{\mu e}&\simeq
\frac{1}{2}\bigg|\epsilon_{13}+e^{i\delta}\big(\epsilon_{12}+i\,\epsilon_\alpha c_{12} s_{12} \big)  \bigg|^2\simeq
\frac{1}{8}\epsilon_\alpha^2 \sin^2 2\theta_{12} - \frac{\epsilon_\alpha \epsilon_{13}}{2} \sin 2\theta_{12} s_\delta + \frac{\epsilon_{12}^2+\epsilon_{13}^2}{2}+\epsilon_{12} \epsilon_{13} c_\delta,
\nonumber \\ 
P_{\mu\tau}&\simeq
\bigg|\epsilon_{23}+\frac{i}{2}\left(\epsilon_\alpha c_{12}^2-\epsilon_\beta \right)\bigg|^2=
\epsilon_{23}^2 + \frac{1}{4} (\epsilon_\alpha c_{12}^2 - \epsilon_\beta)^2.
\label{eq:zero}
\end{align} 
As already inferred in the case of two flavors, the RG effects at zero baseline appear at $\mathcal{O}(\epsilon^2)$ which is clear from 
\cref{eq:zeroe,eq:zeromu,eq:zero}. 
Moreover, it is trivial to show that all asymmetries $P_{\alpha \beta}-P_{\bar\alpha \bar\beta}$ are exactly zero at zero baseline (without any approximations).

For a finite baseline, even approximate expressions for the oscillation probabilities are rather lengthy (full expressions may be found in \cite{Anoka:2005sg}) so here we focus on asymmetries. We keep terms linear in $\epsilon$ and up to order $s_{13}^2$ or $\Delta m^2_{21}/\Delta m^2_{31}$, assuming that the oscillation phase is near the atmospheric maximum. Moreover, for terms that are order $\epsilon$ we only keep terms that are at most linear in $s_{13}$ or $\Delta m^2_{21}/\Delta m^2_{31}$.
The muon neutrino disappearance asymmetry in vacuum is, for example,
\begin{align}
P_{\mu\mu}-P_{\bar{\mu}\bar{\mu}}\simeq &\left\{-(\epsilon_\alpha c^2_{12} -\epsilon_\beta)\sin^2 2\theta_{23}+8\epsilon_{12}c_{13}^2s_{13}c_{23}s_{23}^3s_\delta - \epsilon_\delta s_{23}^4\sin^22\theta_{13}\right\}\, 
  \sin\Delta_{31}-\nonumber\\
  & -\epsilon_{13}\sin2\theta_{12}\sin2\theta_{23}s_\delta\left(1+s_{23}^2\cos\Delta_{31}\right)\sin\Delta_{21}\,.\label{eq:theta12muon}
\end{align}
We will see later that $\theta_{12}$ typically runs more than other angles and phases in the scenarios we will study, so we call attention to the fact that the $\epsilon_{12}$ term in the asymmetry above is suppressed by $s_{13}$. The dependence on $\epsilon_{\alpha}$ and $\epsilon_{\beta}$ is relatively large.
As already stressed in the two flavor scenario, at finite baseline there are effects already at $\mathcal{O}(\epsilon)$; compare for instance \cref{eq:theta12muon} with \cref{eq:zeroe,eq:zeromu,eq:zero}.

The  electron neutrino disappearance asymmetry is given by
\begin{align}
P_{ee}-P_{\bar{e}\bar{e}}\simeq (\epsilon_\beta-\epsilon_\delta) \sin^2 2\theta_{13} \sin\Delta_{31} - \epsilon_\alpha \left( s^2_{12} \sin^2 2\theta_{13} \sin \Delta_{31} - \sin^2 2\theta_{12} \sin \Delta_{21} \right),
\label{eq:finiteLe}
\end{align}
This expression does not depend on $\epsilon_{12}$, $\epsilon_{13}$ and $\epsilon_{23}$ to order $s_{13}^2$ or $\Delta_{21}$. The formulae for $P_{ee}$ and $P_{\bar{e}\bar{e}}$, however, do contain those: the effects cancel in the difference $P_{ee}-P_{\bar{e}\bar{e}}$. Hence, by studying differences between electron neutrino and electron antineutrino disappearances, one can access RG induced effects on phases.
As discussed, the \textit{apparent} violation of CPT symmetry can be seen from \cref{eq:theta12muon,eq:finiteLe}, since the differences between neutrino and antineutrino disappearance are in principle not zero. These differences are CP-odd, as they change sign under the reversal of the signs of all the phases.

The difference between electron neutrino and antineutrino appearance probabilities in a muon (anti)neutrino ``beam'' is somewhat lengthy. Concentrating on the dominant terms up to order $s_{13}$ and dropping $\Delta_{21}$ terms multiplied by new physics contributions, we obtain
\begin{align}
P_{\mu e}-P_{\bar{\mu}\bar{e}} \simeq 
-8 J \Delta_{21} \sin^2\left(\frac{\Delta_{31}}{2}\right) \left[1+\left(2 \frac{\epsilon_{12}}{\sin 2 \theta_{12}} +\epsilon_\alpha \frac{c_\delta}{s_\delta}\right)\frac{\text{cot}(\Delta_{31}/2)}{ \Delta_{21}} \right],
\label{eq:mue}
\end{align}
where $J=c_{13}^2s_{13}c_{12}s_{12}c_{23}s_{23}\sin\delta$ is the Jarlskog invariant~\cite{Jarlskog:1985ht, Jarlskog:1985cw}.
At long baseline experiments, where this channel matters the most, such expansion is reasonable and accurate (ignoring the matter effects). 
We see that in \cref{eq:mue} there is also a term that is $\epsilon$-independent; that is the standard CP violating term.  Interestingly, although the terms containing $\epsilon_{12}$ and $\epsilon_{\alpha}$  are enhanced by $\sim\Delta_{31}/\Delta_{21}$, at the peak of the ``atmospheric'' oscillation $\cot\Delta_{31}/2\simeq\cot\pi/2=0$, which suppresses the CP violating effect. This is particularly pronounced for the T2K setup as will be seen in \cref{sec:results}. We also stress that in the $\delta\to 0$ limit, in which there is no standard CP violation in the lepton sector, new RG induced CP violation is still present and nonzero ($\epsilon_\alpha$ term).
In the future, comparing the amount of leptonic CP violation in DUNE and T2HK will allow one to probe this scenario thanks to the different neutrino energy spectra at the two experiments. 

In order to highlight the CP-conserving contribution, the sum of the electron neutrino and electron antineutrino appearance probabilities, for $\Delta_{21}\to 0$, is
\begin{align}\label{eq:mue-sym}
P_{\mu e}+P_{\bar{\mu}\bar{e}}= 2 \sin^2 {2\theta_{13}} s^2_{23}\left[ 1 
		+ 2 \epsilon_{13} \frac{\cos 2\theta_{13}}{\sin2\theta_{13}} 
		- \frac{c_{23}}{s_{13} s_{23}}\left(\epsilon_{12} c_\delta  - \epsilon_\alpha c_{12}s_{12}s_\delta \right)\right]
			\sin^2\left(\frac{\Delta_{31}}{2}\right).
\end{align}
The first term is the dominant component of the standard contribution. 
It is interesting to observe that the effect of new physics is enhanced by $1/s_{13}\sim7$.
A change in $P_{\mu e}+P_{\bar{\mu}\bar{e}}$ can be compensated, in this channel, by shifting the value of $\sin^2\theta_{23}$ or $\sin^22\theta_{13}$. 
Therefore a mismatch between the $\theta_{23}$ value measured in the $\nu_\mu$ disappearance mode versus the $\nu_e$ appearance mode or a mismatch between $\theta_{13}$ values measured at reactor neutrino experiments and beam $\nu_e$ appearance are signatures of our scenario.
From \cref{eq:mue,eq:mue-sym}, we can infer that the current measurements of electron neutrino appearance by T2K~\cite{Abe:2019ffx} and NOvA~\cite{Acero:2019ksn} should already constrain $\epsilon_{12}$, $\epsilon_{13}$ and $\epsilon_\alpha$ to be below, roughly, 10\%.

We summarize the qualitative effects of the running of the mixing matrix on neutrino oscillation phenomenology and provide some of the most promising and direct experimental probes of this scenario below:

\begin{enumerate}[label=(\roman*)]
  \item In general, the mismatch between the production and detection mixing matrices affects all neutrino oscillation channels.
  
  \item The effect of the phase differences, $\alpha$ and $\beta$ (see \cref{{eq:U3}}), is a shift of the solar ($\Delta_{21}$) and atmospheric ($\Delta_{31}$) oscillation phases, respectively.
  
  \item Zero baseline transitions happen at second order in the new physics parameters. Nonetheless, searches for short-baseline oscillations provide  good experimental probes of this scenario, particularly if performed at high neutrino energies, which makes the neutrino production and detection scales more distinct.
  
  \item The impact of the running of the mixing matrix on long-baseline oscillation probabilities is first order in the new physics parameters and thus the determination of the same oscillation parameters at different scales and the precise energy dependence of the oscillation probability curves are promising venues for probing this scenario.
  
  \item Mixing-matrix running may lead to \textit{apparent} CPT violation: $P_{\alpha \alpha}-P_{\bar\alpha \bar\alpha}$ is, in general, nonzero for a finite baseline. CPT-symmetry is, of course, still conserved. These asymmetries, or perhaps the ratios  $(P_{\alpha \alpha}-P_{\bar\alpha \bar\alpha})/(P_{\alpha \alpha}+P_{\bar\alpha \bar\alpha})$ could be powerful probes of mixing-matrix running, especially due to possible cancelations of systematic uncertainties.
  
  \item Mixing-matrix running may also affect appearance channels in CP-violating and CP-conserving ways. Long-baseline experiments yielding different neutrino energy ranges, including NOvA, T2K, DUNE, and T2HK, could be sensitive to this scenario. Two experimental signatures stand out in the case of electron appearance: a mismatch between $\theta_{13}$ values measured at reactor and accelerator neutrino experiments, or a disagreement on the $\theta_{23}$ values measured in appearance and disappearance modes in beam neutrino experiments. In fact, current NOvA and T2K data are expected to be already sensitive to new sources of CP violation, potentially constraining $\epsilon_{12}$ and $\epsilon_{\alpha}$ to be below 10\% or so.
\end{enumerate}

\subsection{Matter effects: two and more flavors}
\label{subsec:matter}

Neutrino flavor-evolution is modified in the presence of matter. The Hamiltonian that describes flavor-evolution as a function of the baseline is
\begin{equation}
H = \sum_i \frac{m_i^2}{2E}|\nu_i\rangle\langle\nu_i| + A(L) |\nu_e(Q^2=0)\rangle\langle\nu_e(Q^2=0)|,
\label{eq:hamm}
\end{equation}
where $A(L)=\sqrt{2}G_FN_e(L)$ is the matter potential, $G_F$ is the Fermi constant, and $N_e$ is the electron number-density of the medium as a function of the baseline. The matter potential is a coherent forward scattering phenomenon where the neutrinos interact with the electrons in the medium at zero momentum transfer. Hence, the $\nu_e$ interaction state of interest here is the one at $Q^2=0$. $H$ can be expressed in any basis of the Hilbert space, as usual. Here, there are several tempting ones: the mass-eigenstate basis, the interaction-basis ``at production'', and the interaction-basis ``at detection''. The mass-eigenstate basis is especially useful since it allows one to readily compute the flavor-evolution for arbitrary values of $Q^2_p$ and $Q^2_d$. The fact that $H$ depends on $\nu_e(Q^2=0)$ also induces a natural choice for the complex phases $\tilde{\alpha},\tilde{\beta}$, defined in Eq.~(\ref{eq:U3}): $\tilde{\alpha}(Q^2=0)=\tilde{\beta}(Q^2=0)=0$. 

It is instructive to discuss the case of two flavors and a constant $A(L)=A$, which can be solved analytically. Using Eq.~(\ref{eq:U2flavors}) and defining $\theta_0=\theta(Q^2=0)$ and setting $\tilde\beta(Q^2=0)=0$, the eigenstates of the propagation Hamiltonian are
\begin{eqnarray}
|\nu_{1M}\rangle & = & \cos\omega |\nu_1\rangle - \sin\omega|\nu_2\rangle, \\
|\nu_{2M}\rangle & = & \sin\omega |\nu_1\rangle + \cos\omega|\nu_2\rangle, 
\end{eqnarray}
where 
\begin{eqnarray}
\sin 2\omega & = & \frac{A}{\Delta_M}\sin2\theta_0, \\
\cos2\omega & = & \frac{(\Delta-A\cos2\theta_0)}{\Delta_M}, \\
\Delta_M & = & \left[(\Delta-A\cos2\theta_0)^2+A^2\sin^22\theta_0\right]^{1/2},
\end{eqnarray}
and $\Delta=\Delta m^2/(2E)$. $\Delta_M$ is the difference between the eigenvalues associated to $|\nu_{2M}\rangle$ and $|\nu_{1M}\rangle$; $|\nu_{2M}\rangle$ is associated to the larger eigenvalue when $A>0$. We labelled the ``matter mixing angle'' $\omega$ in order to remind the reader that these states are expressed in the mass-basis, not, as one is most familiar, in the interaction basis. 

It is straightforward but rather lengthy to compute $P_{e\mu}\equiv P(\nu_e(Q^2_p)\to \nu_{\mu}(Q^2_d))$. Given a $|\nu_e(Q^2_p)\rangle$ at $L=0$, the flavor-state-vector at $L$ is
\begin{equation}
|\nu(L)\rangle =  |\nu_{1M}\rangle\langle\nu_{1M}|\nu_e(Q_p^2)\rangle + |\nu_{2M}\rangle\langle\nu_{2M}|\nu_e(Q_p^2)\rangle e^{-i\Delta_ML},
\end{equation}
and hence 
\begin{equation}
P_{e\mu}=\left|
\langle\nu_{\mu}(Q^2_d)|\nu_{1M}\rangle\langle\nu_{1M}|\nu_e(Q_p^2)\rangle + \langle\nu_{\mu}(Q_d^2)|\nu_{2M}\rangle\langle\nu_{2M}|\nu_e(Q_p^2)\rangle e^{-i\Delta_ML}
\right|^2.
\label{eq:Pmat2}
\end{equation}
The Dirac brackets in Eq.~(\ref{eq:Pmat2}) are 
\begin{eqnarray}
\langle\nu_{1M}|\nu_e(Q_p^2)\rangle &=& \cos\theta_p\cos\omega-\sin\theta_p\sin\omega e^{i\tilde\beta_{p}}, \nonumber \\
\langle\nu_{2M}|\nu_e(Q_p^2)\rangle &=& \cos\theta_p\sin\omega + \sin\theta_p\cos\omega e^{i\tilde\beta_{p}}, \nonumber \\
\langle\nu_{1M}|\nu_\mu(Q_d^2)\rangle &=& -\sin\theta_d\cos\omega-\cos\theta_d\sin\omega e^{i\tilde\beta_{d}}, \nonumber \\
\langle\nu_{2M}|\nu_\mu(Q_d^2)\rangle &=& -\sin\theta_d\sin\omega+\cos\theta_d\cos\omega e^{i\tilde\beta_{d}}. 
\label{eq:matrixelements}
\end{eqnarray}
Note that since $P_{ee}+P_{e\mu}=1$, the survival probability can be obtained trivially from the appearance one (Eq.~(\ref{eq:Pmat2}) with help from Eqs.~(\ref{eq:matrixelements})).

For three flavors, one can approach the issue of matter effects following the same steps we outline above for two flavors. We especially highlight the usefulness of performing computations in the mass-eigenstate basis. Analytic results, even if one is willing to make several different simplifying assumptions, are very hard to come by and are not illuminating. In Sec.~\ref{sec:results}, we compute oscillation probabilities including running effects for different experimental setups. There, matter effects are always included and we do not make use of any approximate expressions; \cref{eq:hamm} is treated numerically. 

%%%%%%%%%%%%%%%%%%
\section{Benchmark models}
\label{sec:models}
\setcounter{equation}{0}
\setcounter{footnote}{0}
%%%%%%%%%%%%%%%%%%
\noindent
In \cref{sec:running} we discussed the concept of RG evolution in the low-energy neutrino sector without specifying a complete model. In order to discuss quantitative effects at neutrino experiments, however, it is useful to discuss ultraviolet-complete frameworks.  
We will focus on models that address the neutrino mass puzzle and in which the RG effects come from a ``secluded sector.'' We concentrate on two simple models.

\begin{figure}
\centering
	\includegraphics[scale=0.9]{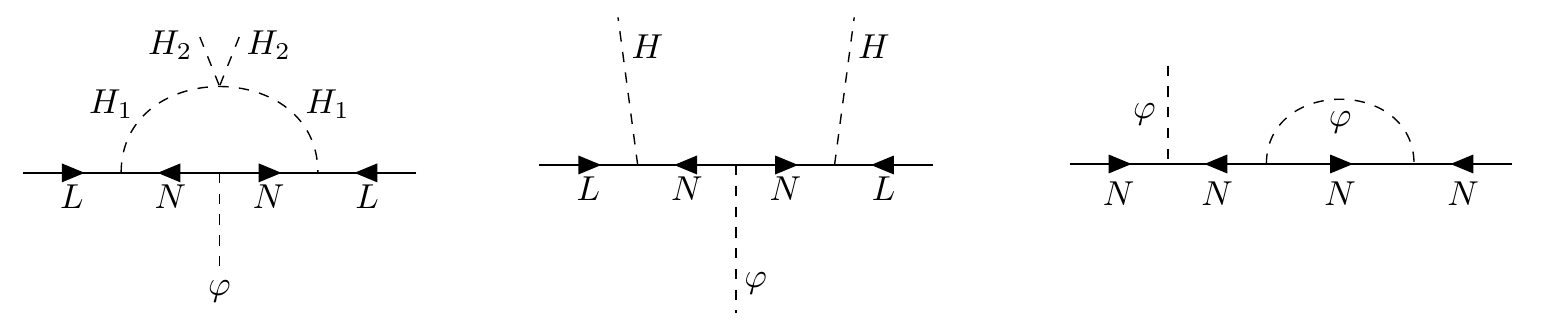} 
	\caption{\emph{Left}: Feynman diagram associated to the generation of neutrino masses in a scotogenic-like neutrino mass model in which $\mathcal{M}\propto Y_N$ (Model 1). \emph{Middle}: Feynman diagram associated to the generation of neutrino masses in an inverse seesaw model in which $\mathcal{M}\propto Y_N^{-1}$ (Model 2). \emph{Right:} Representative Feynman diagram responsible for the running of $Y_N$ in both Model 1 and Model 2. See text for details.}
	\label{fig:feyn}
\end{figure}

{\bf Model 1.} First, we consider a variation of the scotogenic model~\cite{Ma:2006km} with a $U(1)$ lepton number symmetry.
The two Higgs doublets of the scotogenic model, $H_1$ and $H_2$, have zero lepton number but are distinguished by a $\mathbb{Z}_2$ symmetry under which $H_1$ is odd while $H_2$ is even. As in the original scotogenic model, lepton number and the $\mathbb{Z}_2$ symmetries allow the term $\lambda (H_1^\dagger H_2)^2+\text{h.c.}$ in the scalar potential.
The model also comprises three right-handed neutrinos $N_R$, with lepton number $+1$, which are odd under $\mathbb{Z}_2$.
Finally, we add a complex scalar singlet $\varphi$, which is even under $\mathbb{Z}_2$ and has lepton number $-2$. The new physics Lagrangian includes
\begin{equation}
  -\mathcal{L}_{\nu}^{(1)}= \overline{L} Y_\nu \tilde H_1 N_R + \varphi \overline{N_R^c} Y_N N_R + {\rm h.c.},
\end{equation}
where $Y_\nu$ and $Y_N$ are matrices in generation space. 
The scalar potential is such that $\varphi$ and $H_2$ develop vacuum expectation values (vev), while $H_1$ does not.
The active neutrinos acquire Majorana masses, as depicted in the left panel of Fig.~\ref{fig:feyn}. 
The vev of $\varphi$, $v_\varphi\equiv\langle\varphi\rangle$, in particular, breaks lepton number, but the $\mathbb{Z}_2$ symmetry remains unbroken.
Therefore, the active neutrinos do not mix with the gauge-singlets $N_R$. 
Breaking lepton number spontaneously would lead to a massless Goldstone boson, the Majoron~\cite{Chikashige:1980ui}.
In principle, the mass of the Majoron could be made nonzero by soft lepton number breaking terms in the scalar potential, such as $\mu^2\varphi^2$.

Without loss of generality, $Y_N$ can be taken as real, positive, and diagonal. $Y_\nu$ is a generic complex matrix.
The neutrino mass matrix is given by
\begin{equation}\label{eq:masses-1}
  M_\nu^{ij} = \frac{\lambda v_\varphi}{8\sqrt{2}\pi^2}\sum_k\left\{\left(Y_\nu^{ik}Y_\nu^{jk} Y_N^k\right)
      \left[\frac{M_H^2}{2M_H^2- (Y_N^k v_\varphi)^2}\ln\frac{2M_H^2}{(Y_N^k v_\varphi)^2}
      		-\frac{M_A^2}{2M_A^2- (Y_N^k v_\varphi)^2}\ln\frac{2M_A^2}{(Y_N^k v_\varphi)^2}\right]\right\},
\end{equation}
where $M_{H,A}$ are the masses of the inert neutral scalar and pseudoscalar, respectively.
Here, we are interested in light $\varphi$ and $N_R$, and thus $v_\varphi \ll v\equiv 246$~GeV. In the limit of small $N$ masses, i.e., $M_N^i=Y_N^i v_\varphi/\sqrt{2}\ll M_{H,A}$, Eq.~(\ref{eq:masses-1}) reduces to
\begin{equation}\label{eq:nu-mass-1}
    M_\nu \simeq \frac{\lambda v_\varphi}{16\sqrt{2}\pi^2} Y_\nu Y_N Y_\nu^T \ln\frac{M_H^2}{M_A^2}.
\end{equation}
The mixing matrix is defined via diagonalization $M_\nu^{\rm diag}=U^\dagger M_\nu U^*$, where $M_\nu^{\rm diag}$ is diagonal and contains the neutrino masses.
We assume the inert doublet masses to be of order the weak scale so $Y_\nu$ does not run at low energies and can be treated as a constant parameter. $Y_N$, however, is scale dependent for values of the energy scale that are above the mass of the $\varphi,N$.

{\bf Model 2.} The second benchmark model we will focus on is a version of the type-I seesaw mechanism \cite{Minkowski:1977sc, GellMann:1980vs, Yanagida:1979as, Glashow:1979nm, Mohapatra:1979ia, Schechter:1980gr} where the right-handed neutrino Majorana masses arise from the (possibly explicit) breaking of lepton-number in the scalar potential. The model includes 3 SM singlet fermions $N_R$ with lepton number $1$, and a singlet scalar $\varphi$ with lepton number $-2$. The Yukawa Lagrangian reads
\begin{equation}
  -\mathcal{L}_{\nu}^{(2)}= \overline{L} Y_\nu \tilde H N_R + \varphi \overline{N_R^c} Y_N N_R + {\rm h.c.},
\end{equation}
where $H$ is the SM Higgs boson. The vev of $\varphi$ breaks lepton number.\footnote{As in Model 1, this would predict a Majoron. An active-neutrino--Majoron coupling would be induced. It, however, is doubly suppressed by $\nu-N$ mixing and $m_\nu/v_\varphi$, and could be very small (easily of order $10^{-9}$) for the right-handed neutrino masses of interest. Hence the model is expected to be experimentally safe. Again, the Majoron could be given a nonzero mass via soft breaking terms in the scalar potential.}
The diagram generating neutrino masses for this model is presented in the middle panel of Fig.~\ref{fig:feyn}. When $\varphi$ develops a vev, active neutrinos acquire a mass matrix given by
\begin{equation}
  M_\nu = \frac{\sqrt{2}v^2}{4v_\varphi}Y_\nu (Y_N)^{-1} Y_\nu^T.
  \label{eq:nu-mass-2}
\end{equation}
The mixing matrix is defined as before but now the functional dependence of $U$ on $Y_N$ is distinct  from the previous model, changing  the impact  of $Y_N$ running qualitatively with respect to Model 1.

For both models, it is straightforward to compute the scale dependence of $Y_N$.
The right panel of Fig.~\ref{fig:feyn} depicts a representative Feynman diagram that contributes to this running. 
The beta function of $Y_N$ can easily be calculated and yields, for both Model 1 and Model 2, 
\begin{equation}\label{eq:running}
  16 \pi^2 \beta(Y_N)\equiv 16\pi^2\frac{dY_N}{d\ln |Q|}= 4Y_N\left[Y_N^2 + \frac{1}{2}{\rm Tr}(Y_N^2)\right].
\end{equation}
Any relatively large entry in $Y_N$ can lead to a significant running of all $Y_N$ entries, and as a consequence to observable running of the mixing matrix.

$Y_N$ is not directly related to the mixing matrix. In order to connect the running of $Y_N$ and the running of the mixing matrix, we need to specify other Lagrangian parameters, including the Yukawa coupling matrices $Y_{\nu}$ (for both models). To achieve that, we make use of the  Casas-Ibarra parametrization \cite{Casas:2001sr}, which relates the Lagrangian parameters to the running leptonic mixing matrix and neutrino masses:
\begin{align}\label{eq:casas-ibarra}
Y_\nu=\frac{1}{\sqrt{C}}\,U^\dagger(Q_p^2)\,  \sqrt{\text{diag}\left(m_{\nu_1},m_{\nu_2},m_{\nu_3}\right)}\,R\, Y_N^{-x/2}\,,
\end{align}
where $C$ are the vev-dependent prefactors in~\cref{eq:nu-mass-1,eq:nu-mass-2} for Model 1, when $x=1$, and Model 2, when $x=-1$, respectively. $R$ is, in general, a complex orthogonal matrix and $m_{\nu_i}$ in the formula are masses at $Q^2=m_\nu^2$ scale instead of running masses. \newtext{Let us stress that $Y_\nu$ matrix is not running across considered scales since the scalar field which participates in such interaction is integrated out.} In particular, we wish to stress that \cref{eq:casas-ibarra} is evaluated only once, at the scale corresponding to the production, and in particular before any running has been conducted.

In the next section, when computing new-physics effects in oscillation experiments, we have the freedom to choose values for the elements of $R$ and the constants that make up $C$. We will assume $R$ to be real and parameterized by three Euler-like rotation angles $\xi_1$ in 1-2 plane, $\xi_2$ in 2-3 plane and $\xi_3$ in 1-3 plane. The results we present below are not qualitatively different if we are to assume $\xi_1,\xi_2,\xi_3$ to be complex angles with magnitudes of order $1$.

%%%%%%%%%%%%%%%%%%
\section{Impact on neutrino oscillation experiments}
\label{sec:results}
\setcounter{equation}{0}
%%%%%%%%%%%%%%%%%%
\noindent
Given the two concrete models outlined above, 
we proceed with the analysis of the impact of the running of the mixing matrix elements at various neutrino experiments. Results for the two different models turn out to be qualitatively similar and the results presented below all correspond to Model 1. We take this opportunity to stress one ``advantage'' of Model 1, namely the absence of mixing between active and sterile neutrinos. This makes it easier to satisfy experimental constraints on the mixing of the active neutrinos with the relatively-light sterile neutrinos.

In \cref{subsec:long}, we discuss the signatures of the scenario at long-baseline experiments such as T2K \cite{Abe:2018wpn,Abe:2019vii,Abe:2019ffx} and NOvA \cite{Adamson:2017gxd,NOvA:2018gge,Acero:2019ksn} and then we confront  these findings with the bounds from the short-baseline experiments  NOMAD \cite{Vannucci:2014wna,Astier:2001yj,Astier:2003gs}, ICARUS \cite{Antonello:2012pq}, CHARM-II \cite{charm2}, and NuTeV \cite{Naples:1998va,Avvakumov:2002jj}. In \cref{subsec:IC}, we focus on how this phenomenon impacts the flavor composition of astrophysical neutrinos measured by IceCube \cite{Aartsen:2013jdh,Aartsen:2014gkd,Abbasi:2020jmh}.

\subsection{Long-Baseline Oscillation Experiments}
\label{subsec:long}
\noindent
We assume that the masses of the new-physics particles are of order the pion mass so RG running of the mixing parameters is only relevant for $Q^2$ values larger than (100~MeV)$^2$. For lower values of $Q^2$, one can treat the mixing parameters as constant. The reason is we are mostly interested in neutrinos produced at values of $Q_p^2\le m_{\pi}^2$, since all beam neutrinos are predominantly produced in pion decay, and hence, at $Q_p^2$, the mixing matrix is the same for all experimental setups. Furthermore, for reactor and solar (anti)neutrino experiments, both $Q^2_p$ and $Q^2_d$ are less than the pion mass-squared. Hence, we do not need to worry about running effects when it comes to extracting the current best-fit values of most mixing parameters, as we discuss in more detail below.  

Different values for $Y_\nu$ at $Q^2_p$ are generated using Eq.~(\ref{eq:casas-ibarra}) and the following:
\begin{itemize}
\item We fix $\sin^2\theta_{12}(Q_p^2)=0.310$, $\sin^2\theta_{13}(Q_p^2)=0.022$, $\Delta m_{21}^2=7.53\cdot 10^{-5}\,\text{eV}^2$;
\item  We choose the atmospheric parameters $\theta_{23}(Q_p^2)$ and $\Delta m_{31}^2(Q_p^2)$ at random, with a flat prior on their respective $3\sigma$ currently-allowed regions according to NuFIT \cite{Esteban:2020cvm}.
The reason for this choice is that atmospheric parameters are measured using experimental setups where $Q^2_d$ is larger than $m_{\pi}^2$, hence we allow for a relatively large range of values at $Q^2_p$;
 \item  We choose the CP-odd phases $\tilde{\alpha}(Q_p^2), \tilde{\beta}(Q_p^2)$ and $\delta(Q_p^2)$ at random, with a flat prior, from their full allowed physical ranges; 
\item We fix the value of the lightest neutrino mass and the neutrino mass ordering. We will show results for two values: $0.05$~eV, marginally consistent with cosmological bounds on the sum of the active neutrino masses~\cite{Aghanim:2018eyx},
and $0.01$~eV. We will also show results for both normal ordering (NO) and inverted ordering (IO).
Note that, quantitatively, effects depend considerably on the mass ordering and the lightest neutrino mass. It is well known, for example, that the RG effects are strongest for quasidegenerate masses~\cite{Antusch:2005gp};
\item We choose the angles that parameterize the orthogonal $R$ matrix at random, with a flat prior, from their full allowed physical ranges;
\item We pick the Yukawa matrix $Y_N$, at $Q_p^2$, to be $\text{diag}(0.2,0.5,0.7)$. We have checked that the results of the scans are qualitatively independent from the choice of $Y_N(Q^2_P)$ as long as the couplings are non-degenerate and  large enough to induce significant RG running.
\end{itemize}
As an aside, we have checked that the aforementioned choices for the couplings do not lead to the appearance of Landau poles below 10~TeV.

For each mass matrix defined at $Q^2_p$ as described above, we solve \cref{eq:running} numerically and compute the mixing matrix at the different relevant values of $Q^2_d$. With that information, we compute the oscillation probabilities numerically, as discussed in detail in Sec.~\ref{sec:impact}, including  matter effects. Throughout, we will use these randomly generated scenarios to discuss the reach of RG-running effects. A complete scan of the parameter space is not practically feasible given its dimensionality.

At detection, neutrinos interact mostly with the nucleons in the detectors through $t$-channel vector-boson exchange;
the associated $Q^2$ can take any value in a continuous interval.\footnote{This is to be contrasted with what happens in $s$-channel scattering, where the $Q^2$ is defined by the incoming neutrino energy and the mass or energy fraction of the target particle.}
In order to estimate the RG effects accurately, $Q^2_d$ values should be extracted in an event-by-event basis.
Our goals in this manuscript, however, are to illustrate the effects of the running of the mixing matrix in simple models that explain nonzero neutrino masses and to demonstrate that these can be observed in neutrino oscillation experiments. With this in mind, we take a simplified approach that should prove to be a good approximation statistically.
In $2\to2$ scattering kinematics (see e.g. Ref.~\cite{PDG}) the minimal and maximal values of $t=-Q^2$ are fixed, and thus we associate the mean value of the two to $-Q^2_d$.
This yields $Q_d^2= (2 m_N E^2)/(2 E+m_N)$, where $m_N$ is the nucleon mass and $E$  is the neutrino energy.
We take $E$ to be the average neutrino energy in short-baseline experiments while for T2K and NOvA we study the impact of the running for the peak of their respective energy spectra. \cref{tab:1} lists the relevant energies and corresponding $\sqrt{Q_d^2}$ values for all experiments considered here.

%%%%%%%%%%%%%%%%%%%%%%%%%%%%%%%%%%%%%%%%%%%%%%%%%%%%%%%%%%%%%%%
\begin{table}[]
{\color{black}{
{\newcommand{\mr}[3]{\multirow{#1}{#2}{#3}}
\large{
\begin{tabular}{| l |c|c|c|c|}
\hline
Experiment & $E$ (GeV)  & $\sqrt{Q_d^2}$ (GeV) & ~~channel~~ & ~~constraint~~ \\ \hline\hline
T2K \cite{Abe:2018wpn,Abe:2019vii,Abe:2019ffx} & 0.6  & 0.56 & -- &  -- \\ \hline
NOvA \cite{Adamson:2017gxd,NOvA:2018gge,Acero:2019ksn} & 2.1  & 1.27 & -- & --  \\ \hline
ICARUS \cite{Antonello:2012pq} & 17  & 3.94 & $\nu_\mu\to\nu_e$ &  $3.4\times 10^{-3}$ \\ \hline
CHARM-II \cite{charm2} & 24  & 4.70 & $\nu_\mu\to\nu_e$  & $2.8\times 10^{-3}$  \\ \hline
\mr{2}{*}{NOMAD \cite{Vannucci:2014wna,Astier:2001yj,Astier:2003gs}} & \mr{2}{*}{47.5}  & \mr{2}{*}{6.64} & $\nu_\mu\to\nu_e$ & $7.4\times 10^{-3}$  \\ 
													    &   &  & $\nu_\mu\to\nu_\tau$ &  $1.63\times 10^{-4}$ \\ \hline
\mr{3}{*}{NuTeV \cite{Naples:1998va,Avvakumov:2002jj}} & \mr{3}{*}{250}  & \mr{3}{*}{15.30} & $\nu_\mu\to\nu_e$ &  $5.5\times 10^{-4}$ \\ 
										  &   &  & $\nu_e\to\nu_\tau$ & $0.1$    \\
										  &   &  & $\nu_\mu\to\nu_\tau$ &   $9\times 10^{-3}$  \\ \hline
\end{tabular}
}}
\caption{Reference values of neutrino energy and $\sqrt{Q_d^2}$ for all experiments considered here. For T2K and NOvA we take the peak of the neutrino spectra, while for all others (the short-baseline experiments) we use the average neutrino energy. The last two columns summarize the short-baseline constraints imposed (not applicable for T2K and NOvA). See text for details.}
\label{tab:1}
}}
\end{table}
%%%%%%%%%%%%%%%%%%%%%%%%%%%%%%%%%%%%%%%%%%%%%%%%%%%%%%%%%%%%%%

\begin{figure}[t]
	\centering
	\includegraphics[width=0.39\columnwidth]{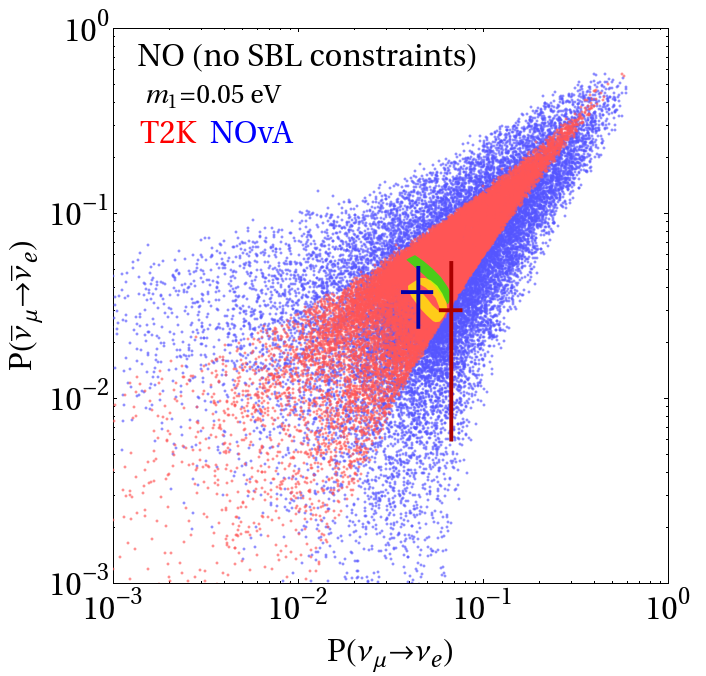} 
	\includegraphics[width=0.39\columnwidth]{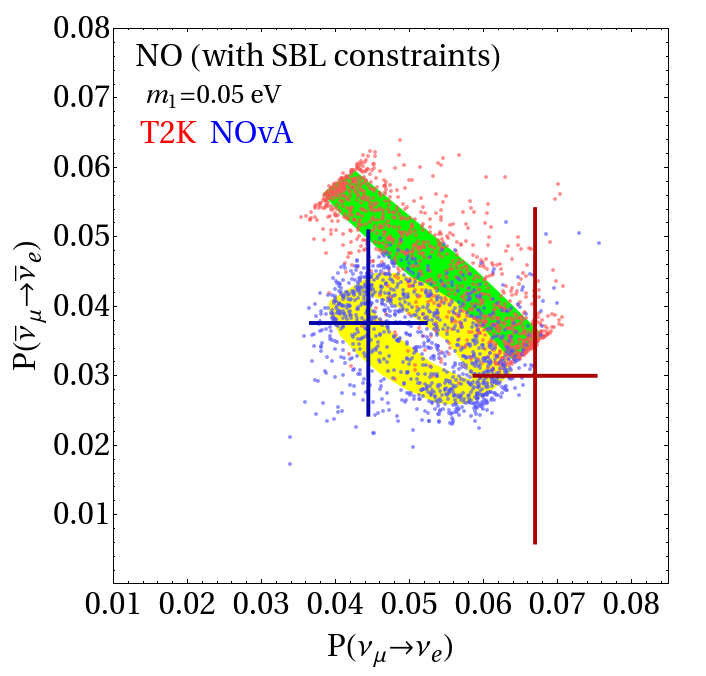} 
	\caption{Bi-probability plots for NO and $m_1=0.05$ eV. \emph{Left}: Red and blue points 
	are for T2K and NOvA, respectively, 
	for the case where RG effects are taken into account. These points are to be compared to the respective standard 1$\sigma$-allowed regions (green for T2K, yellow for NOvA). \emph{Right}: Same as \emph{Left}, except that only points that satisfy the zero baseline constraints are included. See text for details.
	}
	\label{fig:NO5}
\end{figure}

The left panel of \cref{fig:NO5} depicts the bi-probability plot at T2K (red points) and NOvA (blue points) for the NO, assuming the lightest neutrino mass to be $m_1=0.05$ eV, for $30,000$~values of the model parameters, generated following the procedure described above. Note that here we do not take into consideration constraints from other experiments, to which we return momentarily. The panel also depicts the bi-probabilities accessible in the absence of RG running (green for T2K, yellow for NOvA), for values of the atmospheric parameters $\Delta m^2_{31}$ and $\sin^2\theta_{23}$ picked at random (flat prior)  from their respective $1\sigma$ currently-allowed regions, according to NuFIT \cite{Esteban:2020cvm} and for values of $\delta$ also chosen at random from $\delta\in[0,2\pi]$. The red and blue points with error bars represent the results of analyzing T2K and NOvA data, respectively, adapted from Ref.~\cite{Kelly:2020fkv}. 
\cref{fig:NO5} reveals that running effects can be very significant; they lead to appearance probabilities that differ from ``no-running'' expectations by more than an order of magnitude. Effects at NOvA are more pronounced because (a) the typical neutrino energies are larger at NOvA and the mixing parameters have more ``room'' to run and (b) at T2K, as discussed around \cref{eq:mue}, $L/E$ values are such that CP-odd effects due to the new physics are suppressed.

As discussed in Sec.~\ref{sec:impact}, RG running leads to zero baseline effects since, in general, $U(Q_p^2)U^\dagger(Q_d^2)\neq \mathbb{1}$. Therefore, experiments with very short-baselines, designed, with the benefit of hindsight, for probing neutrino scattering physics or light sterile-neutrino phenomenology, are sensitive to this type of new physics.
In particular, setups with high average neutrino energy are especially sensitive due to the larger difference between $Q^2_p$ and $Q^2_d$ and hence potentially stronger running effects.

%%%%%%%%%%%%%%%%%%%%%%%%%%%%%%%%%%%%%%%%%%%%%%%%%%%%%%%%%%%%%%%
%%%%%%%%%%%%%%%%%%%%%%%%%%%%%%%%%%%%%%%%%%%%%%%%%%%%%%%%%%%%%%%

We identified several short-baseline experiments that place stringent constraints on the running: NOMAD~\cite{NOMAD:2001xxt, NOMAD:2003mqg} (bounds from CHORUS \cite{CHORUS:1997wxi,CHORUS:2000jlq} are qualitatively comparable and so are the associated neutrino energies), CHARM-II~\cite{CHARMII:1994rnc}, ICARUS~\cite{ICARUS:2013cwr} and NuTeV~\cite{CCFRNuTeV:1998gjj, NuTeV:2002daf}; see \cref{tab:1} for their reference values of energies and momentum transfers. 
All these experiments measured neutrino beams which were primarily composed of muon neutrinos.\footnote{As a simplifying assumption, we treat all beam neutrinos as if they were the product of pion decay in flight.} Electron-neutrino appearance was severely constrained by all of them, while NOMAD and NuTeV also constrained anomalous $\nu_\mu\to\nu_\tau$ transitions. NuTeV, due to its non-negligible beam-$\nu_e$ component, also managed to put a bound on $\nu_e\to\nu_\tau$ appearance. Concrete upper bounds for each appearance channel are listed in \cref{tab:1}. While NOMAD and NuTeV provide the most stringent bounds on appearance probabilities, it is important to include constraints obtained with different neutrino energies and thus the information provided by the other experiments is invaluable.

The right-hand panel of \cref{fig:NO5} depicts the bi-probability plot for T2K (red points) and NOvA (blue points) for the NO, assuming the lightest neutrino mass to be $m_1=0.05$ eV. As before, we revisit the same 30,000 model-points depicted in the left-hand panel of \cref{fig:NO5} but here only keep those points that satisfy the constraints from the short-baseline experiments listed in \cref{tab:1}. Models associated to a very large effect on NOvA and T2K appearance channels were also very likely to violate the short-baseline constraints. From a model-parameter perspective, zero-baseline constraints disfavor regions of the parameter space with $\mathcal{O}(1)$ Yukawa couplings when the neutrino masses are large. Ultimately, the new physics effects consistent with short-baseline constraints are, at T2K and NOvA ``perturbations'' on standard oscillations. Notice, however, that significant deviations are still allowed;  a significant fraction of the points in the right-hand panel of \cref{fig:NO5} lies outside the standard (no running) $1\sigma$ allowed ranges for T2K and NOvA. 

\begin{figure}[t]
	\centering
	\includegraphics[width=0.39\columnwidth]{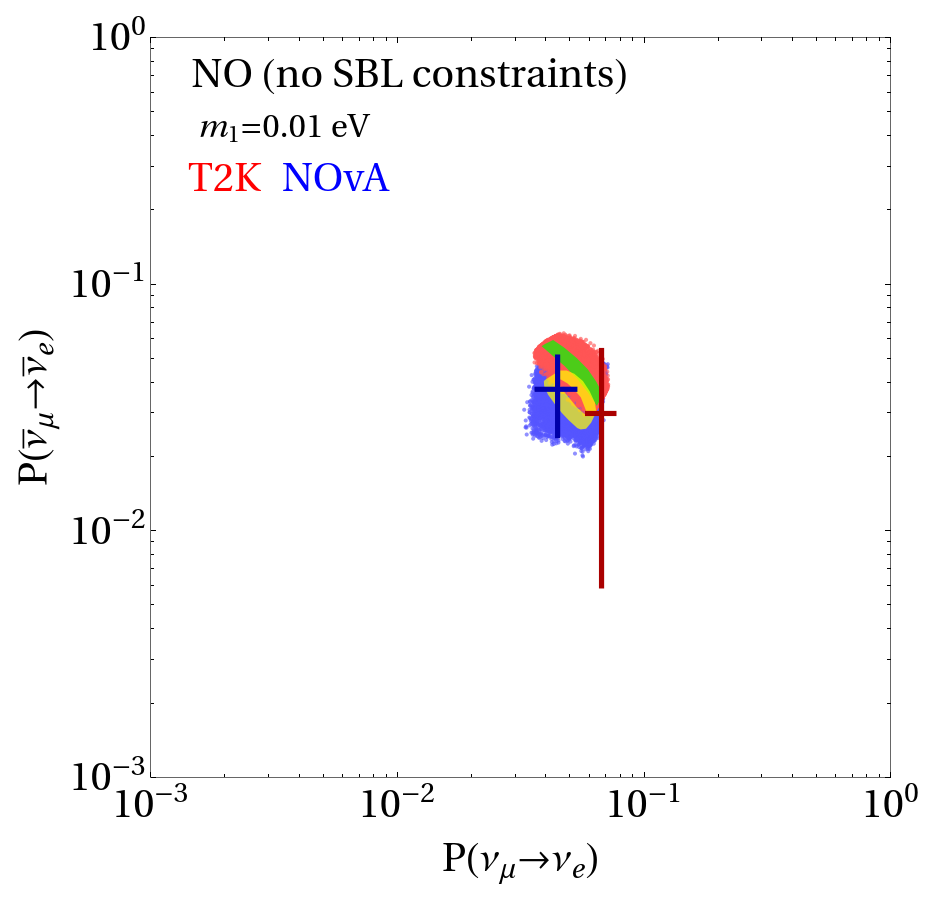} 
	\includegraphics[width=0.39\columnwidth]{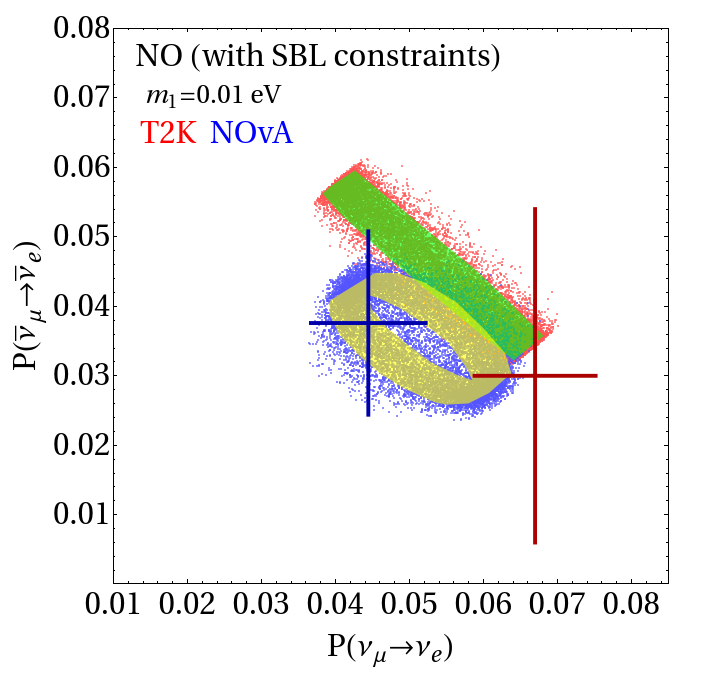} 
	\caption{Same as \cref{fig:NO5} for the case of NO and $m_1=0.01$~eV.
	}
	\label{fig:NO1}
\end{figure}

RG running effects depend strongly on the active neutrino masses and tend to be largest when these are quasi-degenerate. The left panel of \cref{fig:NO1} depicts the bi-probability plot at T2K (red points) and NOvA (blue points) for the NO, this time assuming the lightest neutrino mass to be $m_1=0.01$ eV, for $30,000$~values of the model parameters. The right panel of \cref{fig:NO1} depicts the subset of points that satisfy the short-baseline constraints. 
Comparing  \cref{fig:NO5} with \cref{fig:NO1},  we see that the region of the bi-probability plots accessible to the new physics is larger for quasi-degenerate neutrinos, $m_1=0.05$~eV, when compared to the more hierarchical case, $m_1=0.01$~eV.

\begin{figure}
	\centering
	\includegraphics[width=0.39\columnwidth]{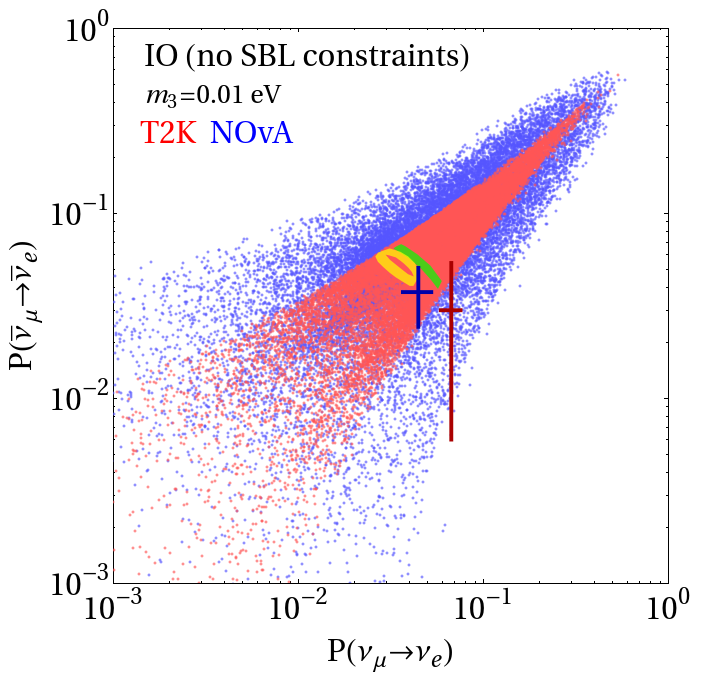} 
	\includegraphics[width=0.39\columnwidth]{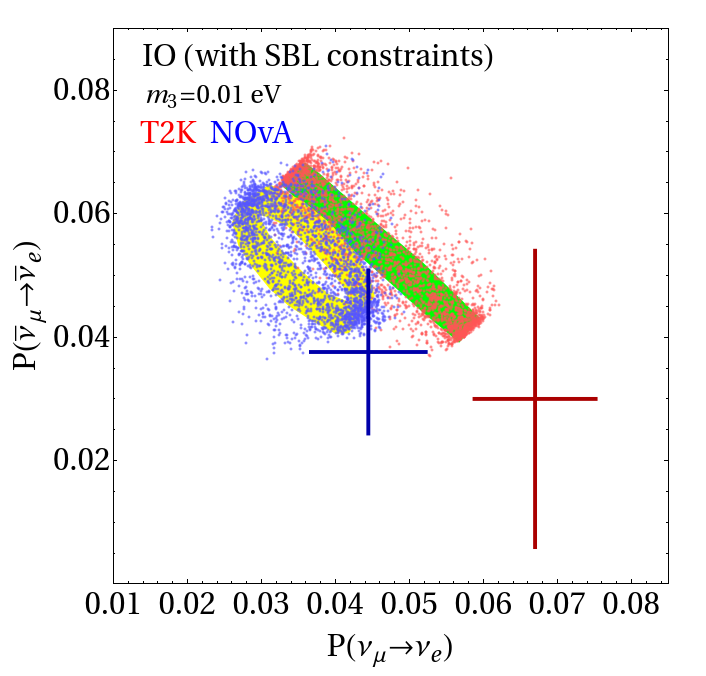} 
	\caption{Same as \cref{fig:NO5} for the case of IO and $m_3=0.01$~eV.
	}
	\label{fig:IO1}
\end{figure}

Similarly, \cref{fig:IO1} depicts the RG effects on T2K and NOvA for the IO and $m_3=0.01$~eV (the lightest neutrino mass). 
Here, even for a relatively light lightest neutrino mass, running effects are comparable to the NO scenario with larger neutrino masses, discussed earlier.
Thus, for the same mass of the lightest neutrino, the RG effects are larger for IO relative to NO.
As expected, the short baseline constraints are very relevant here as well.
%%%%%%%%%%%%%%%%%%%%%%%%%%%%%%%%%%%%%%%%%%%%%%%%%%%%%%%%%%%%%%%
%%%%%%%%%%%%%%%%%%%%%%%%%%%%%%%%%%%%%%%%%%%%%%%%%%%%%%%%%%%%%%%

\begin{figure}
	\centering
	\includegraphics[width=0.39\columnwidth]{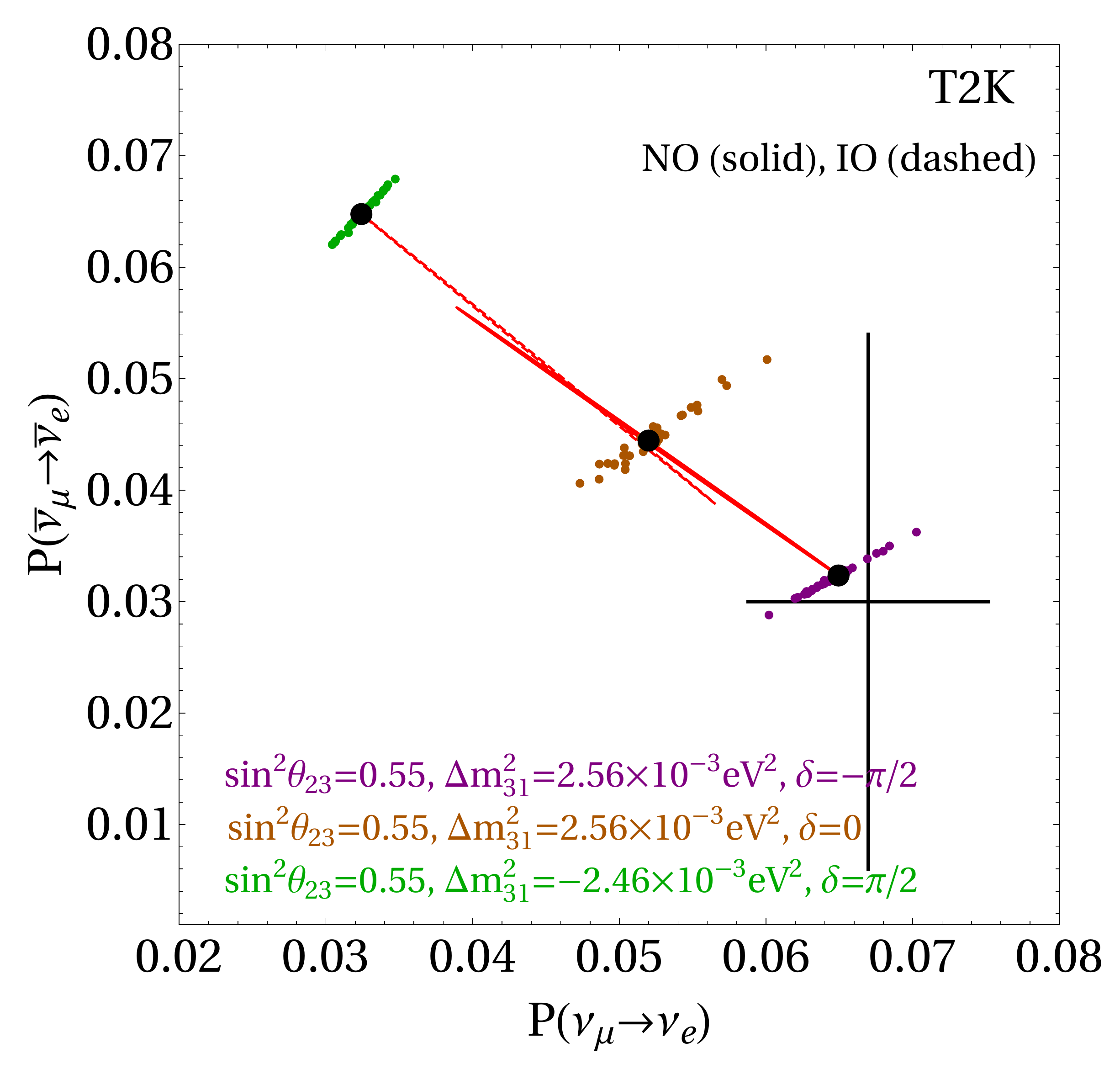} 
	\includegraphics[width=0.39\columnwidth]{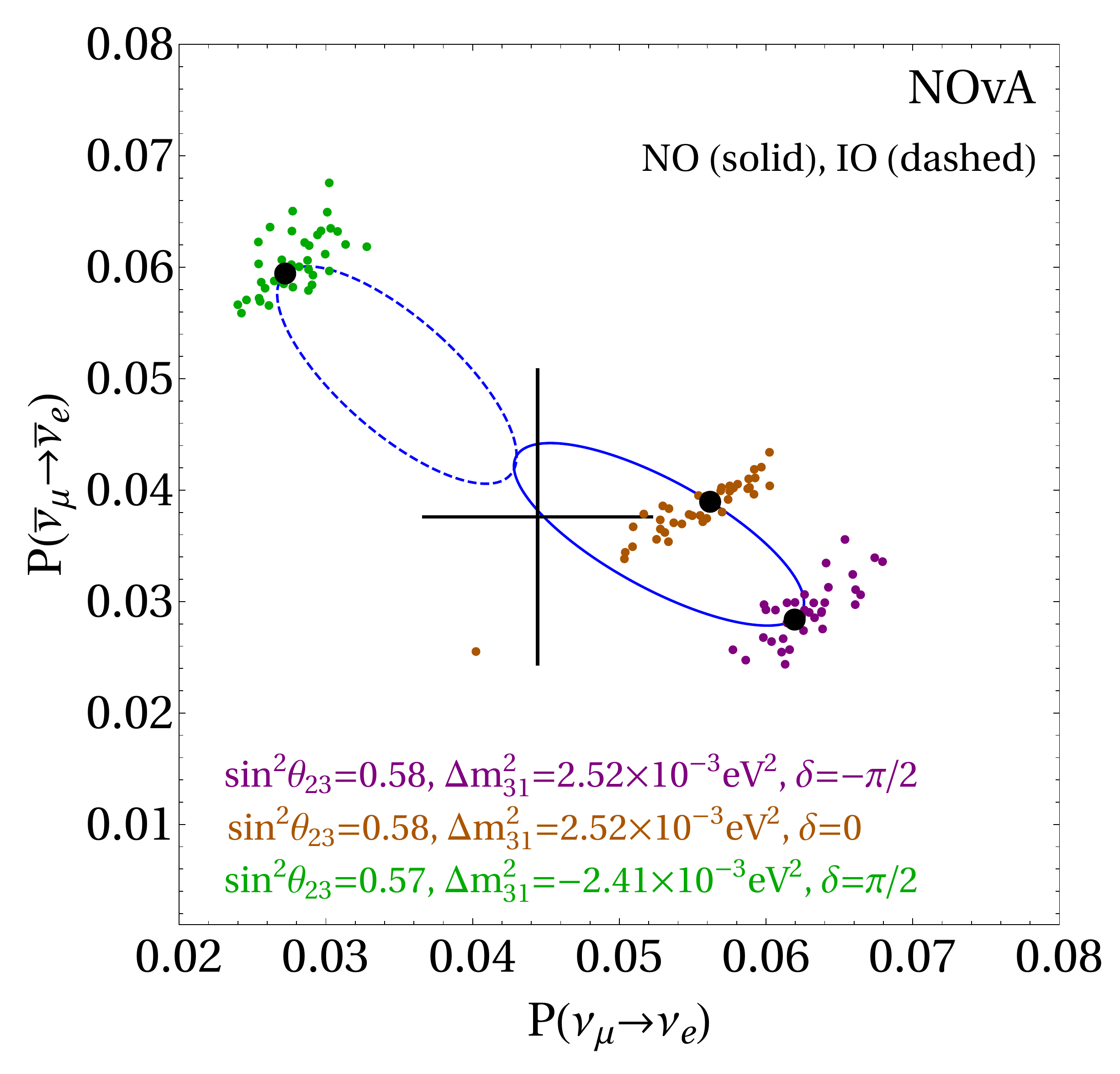}
	\caption{Bi-probability plots -- T2K on the right, NOvA on the left -- for both mass orderings and the lightest mass set to 0.05~eV. The dots indicate the three benchmark points in the standard case. These lie on their respective ellipses, obtained by fixing $\theta_{23}$ and $\Delta m_{31}^2$ and varying $\delta\in [0,2\pi]$. The colored points represent different choices of the new-physics parameters and are indicative of typical RG effects once zero baseline constraints are included. 	}
	\label{fig:individual}
\end{figure}

While \cref{fig:NO5,fig:NO1,fig:IO1} illustrate the general overall reach of RG effect on the appearance channels, it is also useful to understand how RG effects impact oscillations for a specific fixed values of the oscillation parameters at $Q^2_p$.
\cref{fig:individual} depicts the impact of RG running at T2K (left) and NOvA (right), including constraints from the short-baseline experiments, for different fixed values of all oscillation parameters at $Q^2_p$. Modulo rare outliers, the relative new physics effect, that is, the distance from respective black dot, is of order 10\%.
Moreover, the RG effects tend to modify neutrino and antineutrino appearance by similar amounts, while extra contributions to the CP asymmetry are somewhat suppressed.
This can be understood from \cref{eq:mue}, in particular the $\text{cot}(\Delta_{31}/2)$ term that is almost vanishing for T2K at the peak of the energy spectrum.
Similar shifts due to new physics on both neutrino and antineutrino appearances can be mimicked by changing the value of $\sin^2\theta_{23}$.
Therefore, one possible experimental outcome of this scenario could be an apparent inconsistency between the $\sin^2\theta_{23}$ values obtained from the disappearance channel and the $\sin^2\theta_{23}$ from appearance channel.
%%%%%%%%%%%%%%%%%%%%%%%%%%%%%%%%%%%%%%%%%%%%%%%%%%%%%%%%%%%%%%%
%%%%%%%%%%%%%%%%%%%%%%%%%%%%%%%%%%%%%%%%%%%%%%%%%%%%%%%%%%%%%%%

Since RG effects modify oscillation probabilities at T2K and NOvA in slightly different ways, it is interesting to investigate whether they could explain the mild tension between the current T2K and NOvA electron (anti)neutrino appearance data sets. To do this,  we designed a toy $\chi^2$ using the aforementioned T2K and NOvA data points in the bi-probability plane,
\begin{align}
\chi^2=\sum_X \sum_{i=\nu,\bar{\nu}} \left(\frac{D_i^X-T^X_i}{\sigma^X_i} \right)^2\,,
\end{align}
where $X$ denotes T2K or NOvA, $D_i^X$ ($T_i^X$) is the measured (predicted) value of the probability at the experiment $X$ and run mode $i=\nu,\,\bar\nu$, and $\sigma^X_i$ is the error bar read from the plot. Out of the 30,000 generated model-points, we found $\mathcal{O}(10)$ parameter points for NO and $m_1=0.05$~eV which satisfy the short-baseline constraints and provide a slightly better fit to data, though none are statistically significant.
Since the combined fit to T2K and NOvA data leads to a small preference for the inverted ordering~\cite{Kelly:2020fkv}, it is harder to improve the fit in the IO case.
Out of the 30,000 generated model-points for the IO, we found no points which provide a better fit to data compared to standard oscillations.

\begin{figure}[t]
	\centering
	\includegraphics[width=0.39\columnwidth]{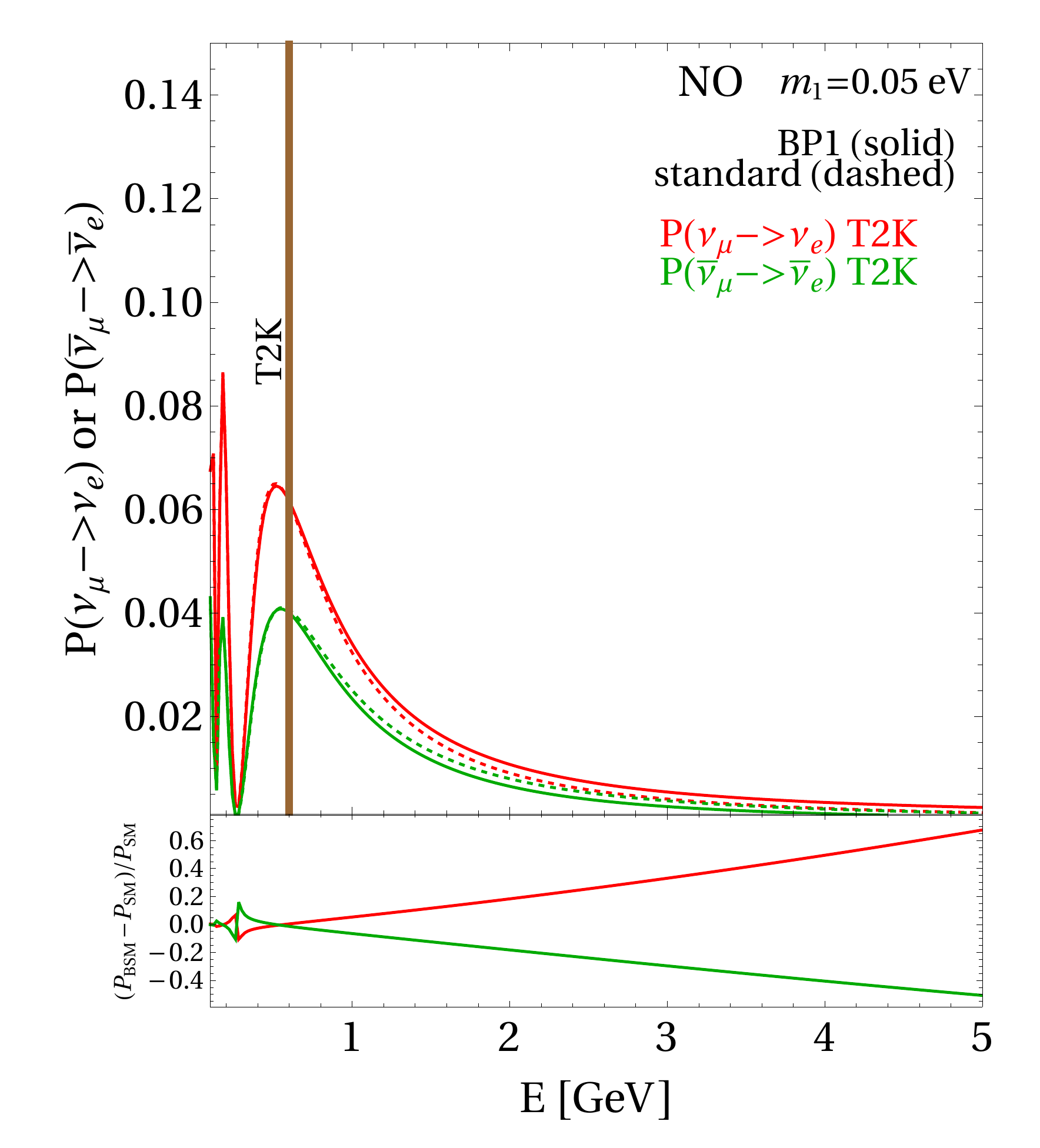} 
	\includegraphics[width=0.39\columnwidth]{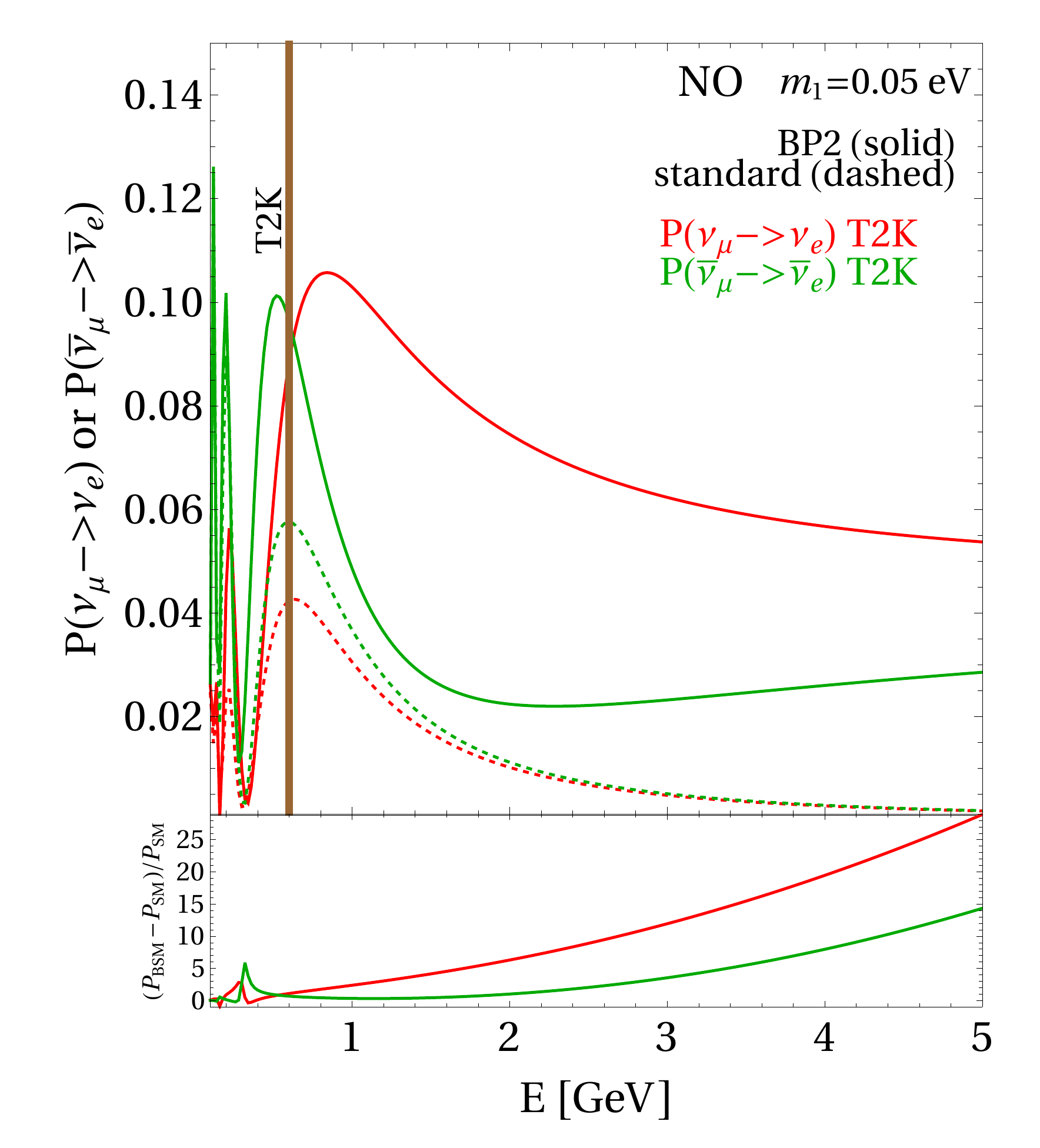}\\ 
	\includegraphics[width=0.39\columnwidth]{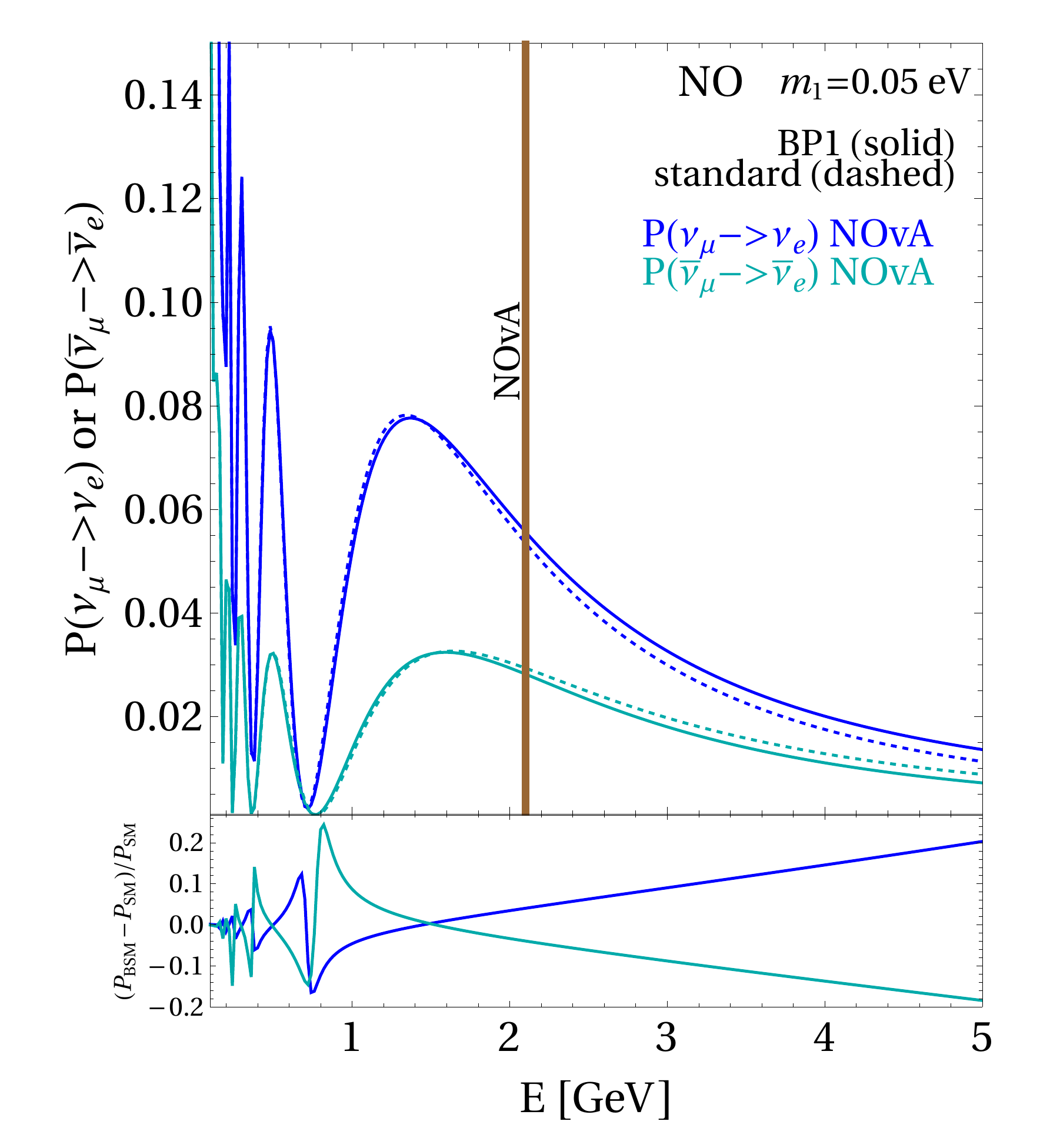} 
	\includegraphics[width=0.39\columnwidth]{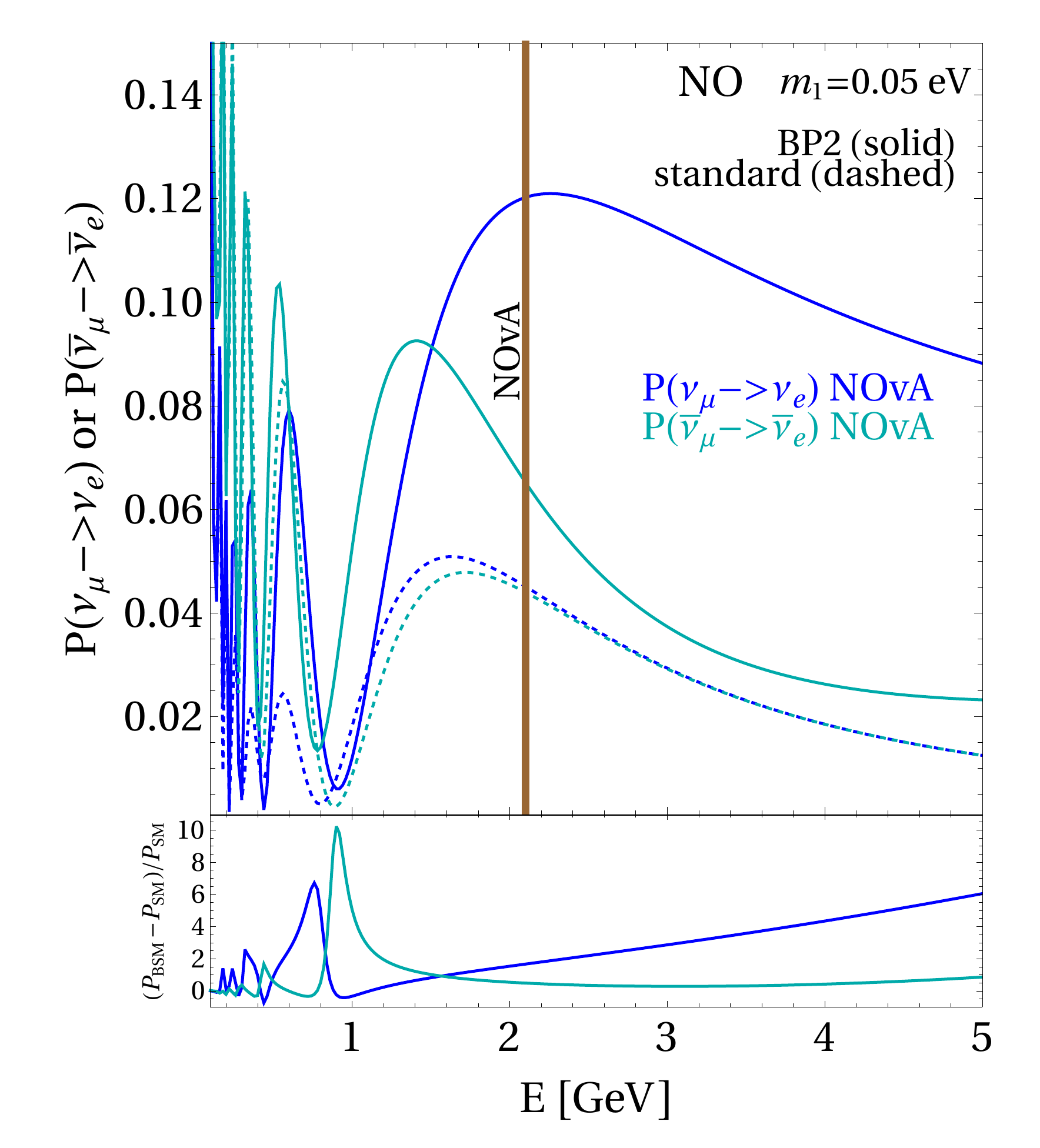}\\ 
	\caption{Oscillation probabilities as a function of the neutrino energy at T2K (upper panels) and NOvA (lower panels) including (solid)  RG-running effects or not (dashed). In the left-hand panels, the new-physics parameters are the ones that provide the best fit we found for the combined T2K and NOvA data. In the right-hand panels, the new-physics parameters are strongly excluded by short-baseline constraints. We assume the NO and the lightest neutrino mass is set to  $0.05$ eV.  The vertical lines indicate the neutrino energies where the T2K and NOvA spectra are largest. The bottom portion of each panel depicts the relative differences between oscillation probabilities with and without RG-running effects.
	}
	\label{fig:probability}
\end{figure}

To further illustrate the RG effects, especially their dependence on the neutrino energy, we depict the electron (anti)neutrino appearance probabilities at T2K (top) and NOvA (bottom), in \cref{fig:probability}, for NO and $m_1=0.05$ eV. 
Red and green curves correspond to the oscillation probabilities for neutrinos and antineutrinos, respectively, at T2K, while curves in shades of blue are constructed by adopting the NOvA far detector baseline. 
The vertical bands indicate the peak of the energy spectrum for each of the two experiments. 
The left-hand panels depict  the oscillation probabilities for the benchmark point in our scan (denoted by BP1)\footnote{
For completeness and reproducibility, we provide the  two benchmark points discussed here. They can be obtained by using \cref{eq:casas-ibarra} with the parameters $(\delta,\,\tilde\alpha,\,\tilde\beta,\,\xi_1,\,\xi_2,\,\xi_3,\,\theta_{23},\,\Delta m_{31}^2/10^{-3}{\rm eV}^2)$ equal to $(3.71,\, 1.57 ,\, 2.37 ,\, 3.45 ,\, 1.51 ,\, 3.00 ,\, 0.88 ,\, 2.437)$ for BP1 and  $(1.18 ,\, 0.24 ,\, 1.64 ,\, 5.48 ,\, 2.076 ,\, 1.85 ,\, 0.86 ,\, 2.525)$ for BP2 (all angles and phases are in radians).} 
 which best fits T2K and NOvA data (solid), compared with the case where the mixing parameters are the same at $Q^2_p$ but there are no running effects (dashed). 
The lower part of each figure depicts the relative difference between these two hypotheses. 
While the differences are small, the relative differences are of order 10\% percent, growing  as the neutrino energy grows due to larger accessible values of the momentum transfer.
The right-hand panels of \cref{fig:probability} depict the case of a different benchmark point, denoted BP2. BP2 is excluded by the short-baseline data. Nonetheless, it serves to illustrate the possible impact of RG effects on oscillation experiments and to highlight the importance of the short-baseline experiments.
%%%%%%%%%%%%%%%%%%%%%%%%%%%%%%%%%%%%%%%%%%%%%%%%%%%%%%%%%%%%%%%
%%%%%%%%%%%%%%%%%%%%%%%%%%%%%%%%%%%%%%%%%%%%%%%%%%%%%%%%%%%%%%%

\begin{figure}[t]
	\centering
	\includegraphics[width=0.39\columnwidth]{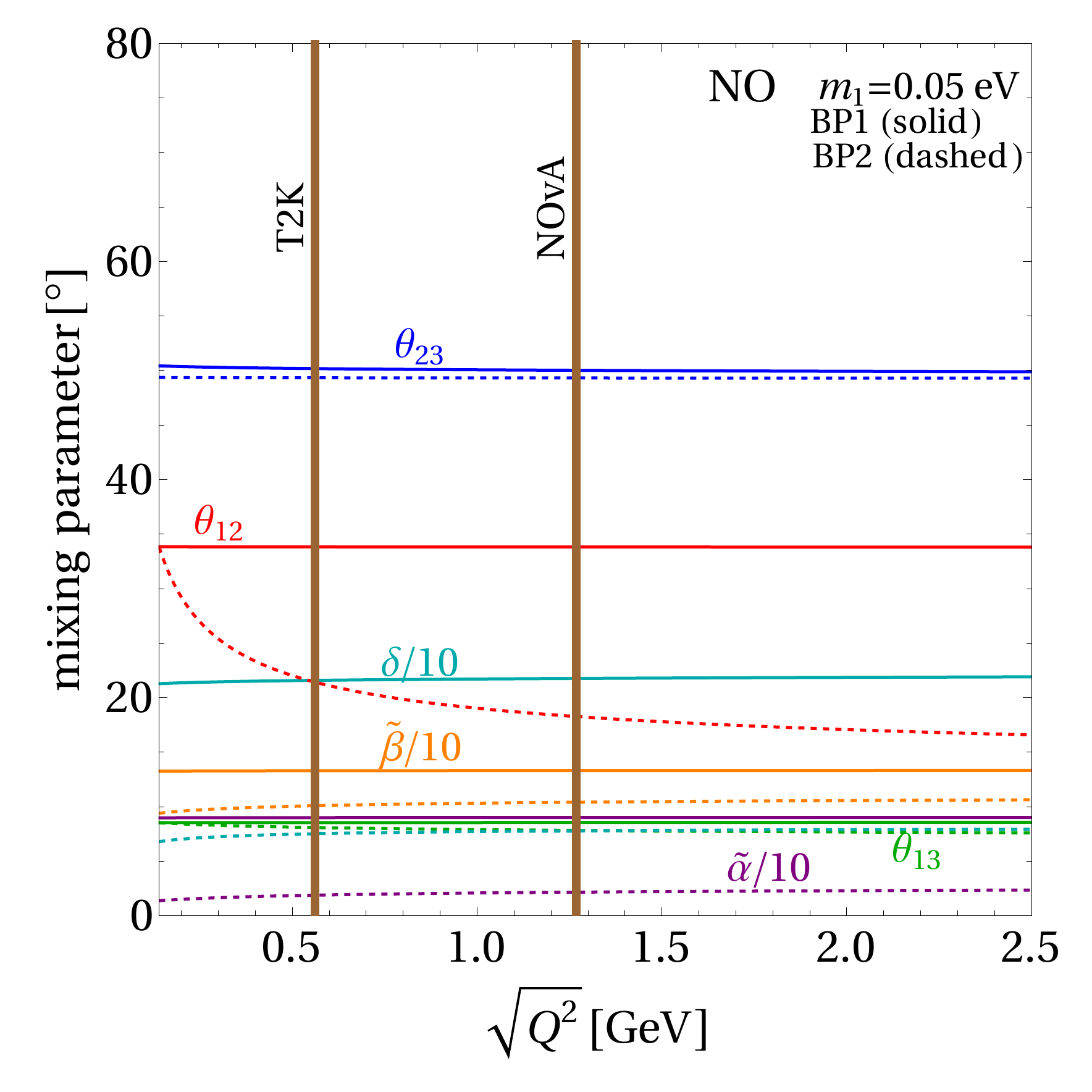} 
	\includegraphics[width=0.39\columnwidth]{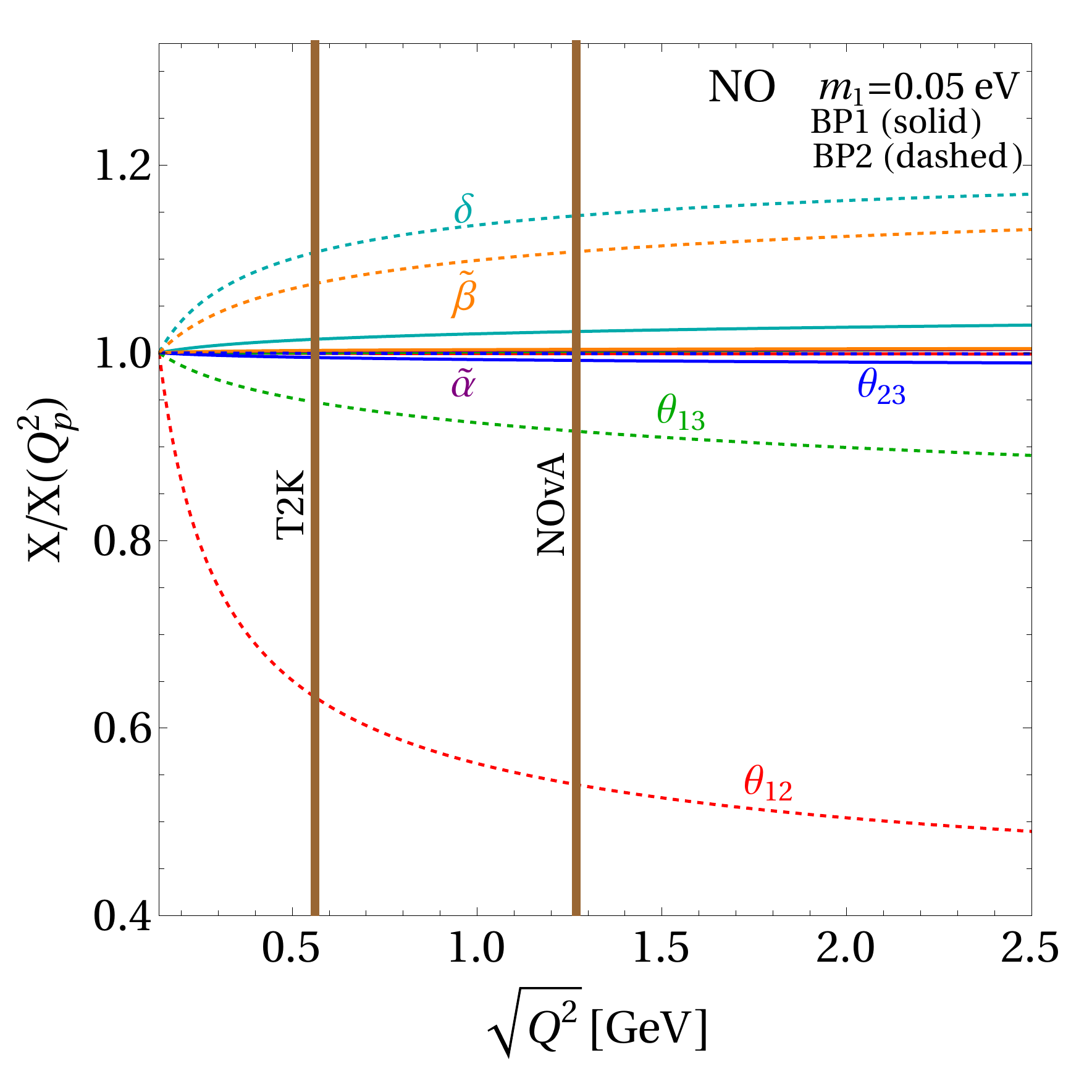} 
	\caption{\emph{Left}: RG evolution of the mixing parameters for the two new-physics parameter points in \cref{fig:probability}. The solid (dashed) lines are for the best-fit (strongly excluded) point. The vertical lines denote the peak-value of $\sqrt{Q^2}$ at T2K and NOvA. \emph{Right}: RG evolution of the mixing parameters, normalized by the respective values at $Q_p^2$.
 }
	\label{fig:running}
\end{figure}

\cref{fig:running} depicts the RG evolution of the relevant mixing parameters for both benchmark points BP1 and BP2.
The left-hand panel depicts the values of the different parameters as a function of $\sqrt{Q^2}$ while  the right-hand panel depicts the ratio between each parameter at $Q^2$ relative to its value at $Q_p^2$. For BP1 (solid), RG evolution yields few percent-level changes in the values of oscillation parameters. 
Running effects are more pronounced in the case of BP2 (dashed). The strongest effects are in running of $\theta_{12}$, which is very strongly impacted by RG effects. 
This large variation in $\theta_{12}$ disturbs all flavor transitions, and induces sizable zero-baseline effects as $U(Q^2_d)U^\dagger(Q^2_p)$ strongly deviates from unity.
Hence, these points in model-space are strongly constrained by both short- and long-baseline experiments. 
In general we find that, when strong RG evolution effects are present, they typically first appear in $\theta_{12}$.
It is well known~\cite{Antusch:2005gp} that the variation of $\theta_{12}$ relative to the other mixing angles $\theta_{13}$ and $\theta_{23}$ is enhanced by the ratio between atmospheric and solar mass-squared differences: $|\Delta \theta_{12}/\Delta\theta_{13}|,|\Delta \theta_{12}/\Delta\theta_{23}|\propto|\Delta m_{31}^2/\Delta m_{21}^2|$.\footnote{In our phenomenological discussions of RG running effects at T2K and NOvA, we did not highlight the fact that $\theta_{12}$-running effects are largest. The reason is that long-baseline experiments (and all other ``high-energy'' experiments, including measurements of the  atmospheric neutrino flux) have limited ability to constrain the solar oscillation parameters since, for the typical energies associated to these setups $E/\Delta m^2$ is much larger than the relevant baselines. While higher precision and more statistics are expected at DUNE and Hyper-Kamiokande, these will also have only limited ability to independently measure $\theta_{12}$.}
%%%%%%%%%%%%%%%%%%%%%%%%%%%%%%%%%%%%%%%%%%%%%%%%%%%%%%%%%%%%%%%
%%%%%%%%%%%%%%%%%%%%%%%%%%%%%%%%%%%%%%%%%%%%%%%%%%%%%%%%%%%%%%%

Finally, since RG effects on short-baseline experiment can be sizable, one may be tempted to search for explanations to the LSND~\cite{Athanassopoulos:1996jb} and/or MiniBooNE~\cite{Aguilar-Arevalo:2018gpe,Aguilar-Arevalo:2020nvw} anomalies using this framework. 
In particular, we found plenty of points in model-space with zero-baseline appearance probability $P_{\nu_\mu\to \nu_e}\simeq 10^{-3}$ at around $E_\nu\simeq 0.4$ GeV, hence qualitatively in agreement with the oscillation interpretation of the MiniBooNE data.
However, once constraints from NOMAD and NuTeV are imposed, such points are ruled out. 
This is mostly because RG effects are stronger at larger neutrino energies.
We will explore the short-baseline phenomenology of this framework and variations thereof in more detail in an upcoming manuscript.

\subsection{Ultra-High-Energy Neutrinos from the Cosmos}
\label{subsec:IC}

Neutrino oscillation experiments are not the only way to search for the RG running of neutrino mixing parameters. In particular, RG effects grow with $Q^2$ and the IceCube experiment has detected neutrinos from extra-galactic astrophysical sources with laboratory energies up to the PeV-scale. This correspond to $\sqrt{Q_d^2}\simeq 10^3$ GeV, far above the corresponding values accessible to terrestrial experiments (see \cref{tab:1}).  Here we argue that precise measurements of the flavor-composition of ultra-high-energy (UHE) neutrinos at IceCube are also sensitive to the new physics effects discussed here.  

In \cref{subsec:long}, we focused on RG effects for $m_\pi^2\le Q^2_d \lesssim (16)^2$~GeV$^2$ and concluded that new-physics effects were a perturbation over the standard expectation, particularly due to constraints from short-baseline experiments. Here, we will modify the RG running conditions and assume that the RG effects only take place for $\sqrt{Q^2}\gg16$~GeV; i.e., the new degrees of freedom have masses of order tens of GeV. This assumption leads to no observable effects at solar-system neutrino experiments but allows for potentially strong effects on measurements of the flavor composition of IceCube's UHE neutrinos.
For the purpose  of our calculations, we postulate that running starts at $\sqrt{Q_p^2}=16$~GeV and the values of all the parameters at  $\sqrt{Q^2}=16$~GeV are consistent with existing constraints from oscillation experiments. It is important to note that UHE neutrinos are produced, for the most part, in the decays of pions, muons or perhaps neutrons so $Q_p^2$ is usually around or below $m_{\pi}^2$, independent from the neutrino energy. Since the running only starts at the mass scale of new particles, calculating the running from $m_\pi^2$ to $Q^2_d$ is  equivalent to setting the production $Q^2_p$ to the mass-scale of new particles. As far as detection at IceCube is concerned, we fix $\sqrt{Q_d^2}=10^3$~GeV, for simplicity.

The IceCube collaboration has released results \cite{Aartsen:2015knd,HESEflavor} on the flavor composition of astrophysical neutrinos, an observable that is also one of the pillars of the forthcoming IceCube-Gen2 upgrade \cite{Shoemaker:2015qul,Aartsen:2020fgd,Song:2020nfh}. 
The initial flavor composition ($\nu_e$:$\nu_\mu$:$\nu_\tau$) of $\mathcal{O}$(PeV) neutrinos (note that, here, we do not distinguish between neutrinos and antineutrinos) from astrophysical neutrino sources is uncertain due to our poor knowledge of the nature of these sources and the mechanism behind neutrino production. We will consider several hypotheses.
If the neutrinos are predominantly produced in the decays of high energy pions and in the decays of the pion-daughter muons, the flavor ratios at production would be  (1:2:0)~\cite{Kashti:2005qa,Winter:2014pya}. 
If, on the other hand, muons from the pion decays lose most of the their energy before decaying (e.g., due to interactions in a dense medium~\cite{Sui:2018bbh}), their decay products are not very energetic and the initial flavor composition would be (0:1:0)~\cite{Waxman:1997ti,Rachen:1998fd}. If, on the other hand, neutron decays are the dominant source of UHE neutrinos, only electron antineutrinos would be produced at the source: (1:0:0)~\cite{Lipari:2007su}. It turns out that neutron decays as the dominant source of UHE neutrinos are disfavored at around the  68\%CL by Icecube data~\cite{Aartsen:2015knd,HESEflavor}. For completeness, one could also consider a pure tau neutrino initial flavor composition (0:0:1)~\cite{Arguelles:2015dca,Bustamante:2015waa,Brdar:2016thq} even if this is not expected to occur in any known astrophysical environment. 

Once produced, the propagation of neutrinos to the Earth is subject to neutrino oscillations. Given the very long baselines, it is safe to treat these neutrinos are incoherent superpositions of mass eigenstates and the oscillation probabilities at Earth are baseline independent.
In the standard three-neutrino paradigm, the probability of producing an astrophysical neutrino of flavor $\alpha$ and detecting it with flavor $\beta$, $\alpha,\beta=e,\mu,\tau$, is given by
\begin{align}
  P_{\nu_\alpha \to \nu_\beta}= P_{\nu_\beta \to \nu_\alpha}
    = \delta_{\alpha\beta}- 2\sum_{k>j} \text{Re} \big[ U_{\alpha k}^* U_{\beta k} U_{\alpha j} U_{\beta j}^* \big]
    = \sum_{j=1}^n \big|U_{\alpha j}\big|^2 \big|U_{\beta j}\big|^2 \,.
  \label{eq:P_standard}
\end{align}
In the presence of mixing-matrix running, these probabilities are \cite{Bustamante:2010bf}
\begin{align}
P_{\nu_\alpha \to \nu_\beta}=
\label {eq:P_new}\sum_{j=1}^3 \big|U_{\alpha j}{(Q_p^2)}\big|^2 \big|U_{\beta j}{(Q_d^2)}\big|^2 \,,
\end{align}
where we again stress that mixing matrix elements evaluated at the $Q_p^2$ and $Q_d^2$ scales are potentially different.
The expected flavor composition at the surface of the Earth is
\begin{align}
X_\beta = \sum_\alpha P_{\nu_\alpha \to \nu_\beta} X_\alpha^{\rm prod}\,,
\end{align}
where $X_\alpha^{\rm prod}$ is the fraction of neutrinos of flavor $\alpha$ at production.

\begin{figure}
	\centering
	\includegraphics[width=0.85\columnwidth]{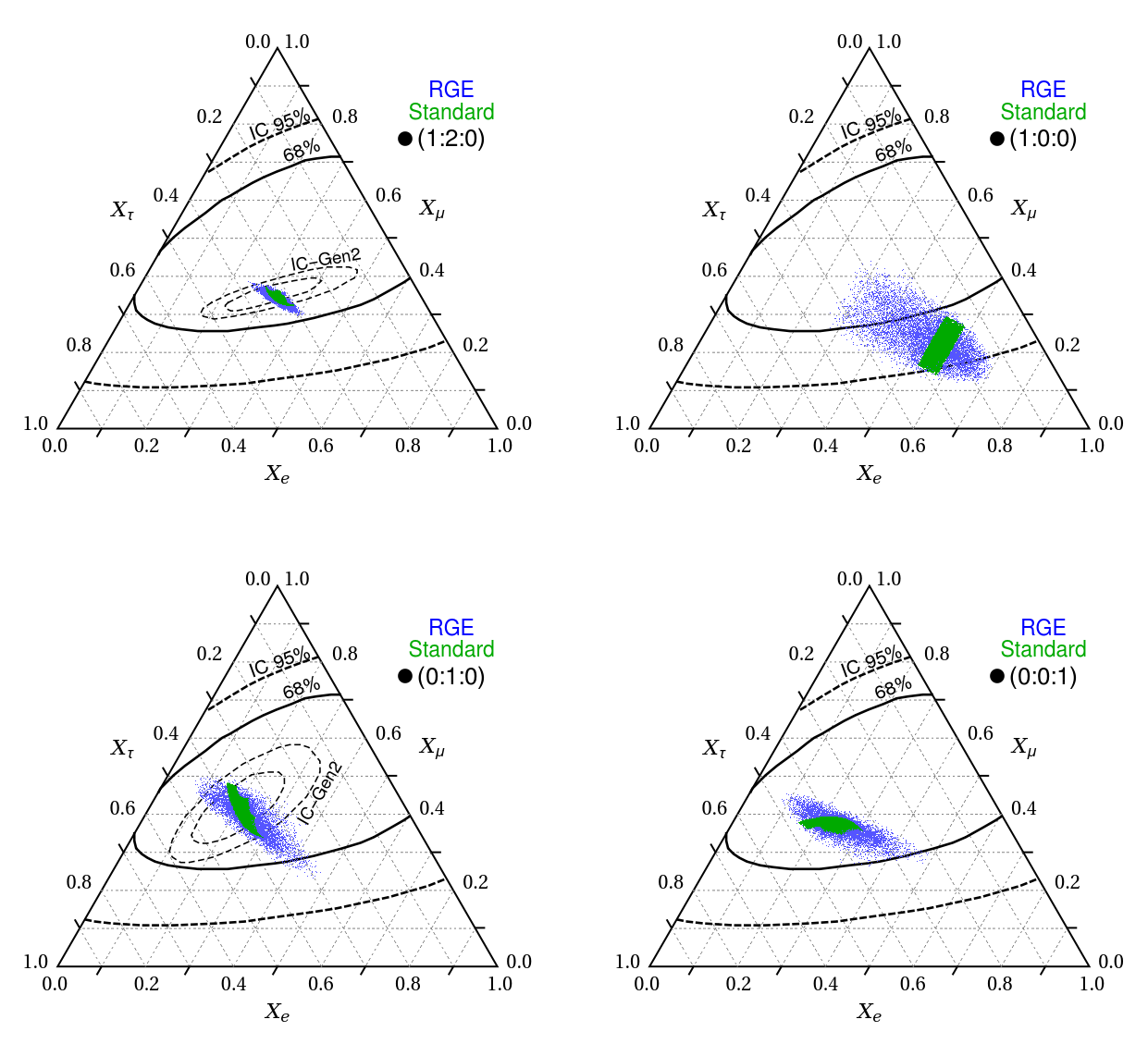} 
	\caption{The relative flavor composition of UHE astrophysical neutrinos at the Earth for different choices of the relative flavor composition at the source. 
	The green region corresponds to expectations from standard three-neutrino oscillations while the blue scatter points represent the potential effects of mixing-matrix running. 
	Present limits from IceCube are shown as thick dashed (68\% C.L.) and thick solid (95\% C.L.) lines. 
	For pion-decay and damped-muon sources (left-top and left-bottom, respectively), we also show IceCube Gen-2 projections as thin dashed lines (68\% and 95\% C.L.). See text for details.}
	\label{fig:IC1}
\end{figure}

\cref{fig:IC1} depicts the flavor compositions in the ternary plots that are often employed for this type of study \cite{article,Learned:1994wg}. 
Each panel corresponds to one of the production scenarios discussed above. 
Green regions indicate the accessible range in the flavor triangle for standard neutrino oscillations, whereas blue scatter points represent different scenarios with RG effects included. 
The parameter-scan strategy is similar to the one discussed in \cref{subsec:long} for both the standard scenario and the different new-physics models.
In all panels, the new-physics effects increase the accessible region of the flavor-triangle when compared to the standard case due to the mismatch between production and detection mixing matrix elements. 
The most striking effect can be seen in the upper-right-hand panel, associated to (1:0:0) (neutron decay source) at production. 
As we discussed, while this type of production is disfavored by present data in the absence of new physics, the situation is different once RG running effects are included; as can be seen in the figure, there are plenty of blue points safely located inside the 68\% CL region.
%%%%%%%%%%%%%%%%%%%%%%%%%%%%%%%%%%%%%%%%%%%%%%%%%%%%%%%%%%%%%%%
%%%%%%%%%%%%%%%%%%%%%%%%%%%%%%%%%%%%%%%%%%%%%%%%%%%%%%%%%%%%%%%

\begin{figure}
	\centering
	\includegraphics[width=0.5\columnwidth]{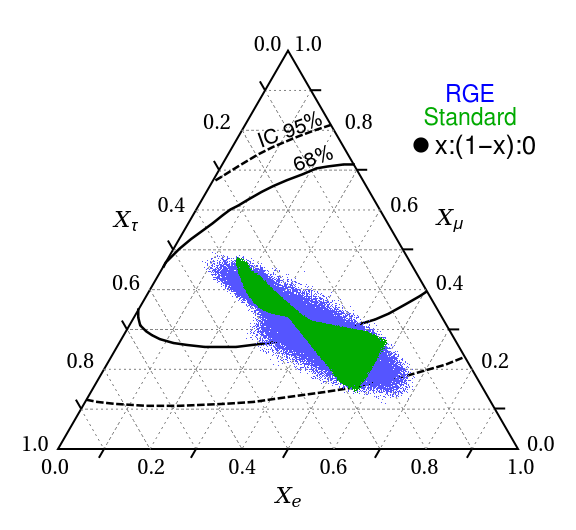} 
	\caption{Same as \cref{fig:IC1}, for all possible flavor-compositions ($x\in[0,1]$), assuming the sources can only produce electron and muon neutrinos.  
	}
	\label{fig:IC2}
\end{figure}

For the preferred (1:2:0) and (0:1:0) scenarios, we also depict, in addition to present constraints, the future projections of IceCube-Gen2~\cite{Aartsen:2020fgd}. 
The region defined by the blue scatter points exceeds those regions significantly. 
Hence, as the efforts in the field of neutrino astrophysics lead to the discovery of point sources and illuminate the neutrino production mechanism, it will be possible to use the flavor-composition observable to probe the presence of RG induced new physics effects with some precision.  Finally, we also consider the more general production scenario in which the flavor composition is a generic mixture of electron and muon neutrinos. \cref{fig:IC2} depicts the flavor triangles for initial flavor ratios given by ($x$:$(1-x)$:$0$), $x \in [0,1]$, where we scan over all possible values of $x\in[0,1]$. As in \cref{fig:IC1}, it is clearly visible that mixing-matrix running may lead to strong effects on the flavor composition at the Earth. We can repeat this exercise for a most-general-source of the type ($x(1-y)$:$(1-x)(1-y)$:$y$), $x,y \in [0,1]$. In this case, we also  find that running effects can land well outside the region one can access in the context of the standard scenario.

We conclude this subsection by summarizing the main effects induced by the running of the mixing matrix:
\begin{itemize}
\item Even if the running of the mixing matrix starts at energy scales higher than those accessible to solar-system neutrino experiments, the impact on the  flavor composition of UHE neutrinos can be quite large.
\item The main effect of the running is to enlarge the set of allowed values for the flavor-ratios at IceCube, for all production mechanism. A signature of this scenario is the measurement of a flavor composition of the UHE neutrino flux that is inconsistent with expectations from standard three-neutrino oscillations.
\end{itemize}

%%%%%%%%%%%%%%%%%%
\section{Conclusions}
\label{sec:conclusion}
\setcounter{equation}{0}
%%%%%%%%%%%%%%%%%%
\noindent
We considered, within the context of simple, ultraviolet-complete models of neutrino masses, the effects of scale-dependent lepton mixing parameters at neutrino oscillation experiments. 
In this framework, the mixing matrix at production and detection may be different, leading to rich and novel phenomenological consequences. We identified several robust experimental signatures of this framework, including:

\begin{itemize}

\item  Apparent mismatches between $\theta_{13}$ measurements at reactor and beam experiments as well as apparent mismatches between $\theta_{23}$ measurements in $\nu_\mu$ disappearance and $\nu_e$ appearance channels. This would be the smoking gun signature of the framework proposed in this paper\,,

\item  new sources of CP-invariance violation\,,

\item zero-baseline flavor transitions $\nu_\mu\to\nu_\tau$, $\nu_\mu\to\nu_e$ and $\nu_e\to\nu_\tau$\,,

\item apparent CPT-invariance violation in the form $P(\nu_\alpha\to\nu_\alpha)\neq P(\bar\nu_\alpha\to\bar\nu_\alpha)$ for  $\alpha=e,\mu,\tau$\,.

\end{itemize}

Taking current experimental constraints from short-baseline experiments into account, we showed that the renormalization group evolution of the mixing parameters can induce observable effects at T2K, NOvA, and future long-baseline neutrino experiments. As a complementary probe to short- and long-baseline experiments, we also scrutinized effects at neutrino telescopes, in particular those related to the flavor composition of ultra-high-energy neutrinos. 

Observable effects of RG running of the leptonic mixing matrix are a potential consequence of new, relatively light degrees of freedom and new neutrino interactions. As we demonstrated, the new interactions may be restricted, mostly, to the neutrino sector and hence are difficult to constrain outside of experiments that involve flavor-resolved neutrino scattering. The effects discussed here are qualitatively different from the more familiar nonstandard neutral-current-like neutrino interactions, which often manifest themselves via modified matter effects or new interactions that impact neutrino production or detection. The latter, in particular, may also be described using different leptonic mixing matrices at neutrino production relative to neutrino detection, but do not share the scale dependency of the RG running effects discussed here. New, light degrees of freedom can also be directly produced in neutrino scattering. These effects are complementary to what we are discussing here, and have been explored, rather recently, in the literature.

%%%%%%%%%%%%%
\section*{ACKNOWLEDGEMENTS}
\noindent
We thank Bill Bardeen and Joachim Kopp for useful discussions.
This work was supported in part by the US Department of Energy (DOE) grant \#de-sc0010143 and in part by the NSF grant PHY-1630782. 
Fermilab is managed by Fermi Research Alliance, LLC (FRA), acting under Contract No. DE-AC02-07CH11359. 
This project has received support from the European Union’s Horizon 2020 research and innovation programme under the
Marie Sk\l{}odowska-Curie grant agreement No 860881-HIDDeN.
%%%%%%%%%%%%%

\bibliographystyle{JHEP}
\bibliography{nu_run}

\end{document}